\PassOptionsToPackage{unicode}{hyperref}
\PassOptionsToPackage{hyphens}{url}
\documentclass[10pt]{article}
\usepackage{amsmath,amssymb}
\usepackage{iftex}
\ifPDFTeX
  \usepackage[T1]{fontenc}
  \usepackage[utf8]{inputenc}
  \usepackage{textcomp} 
\else 
  \usepackage{unicode-math} 
  \defaultfontfeatures{Scale=MatchLowercase}
  \defaultfontfeatures[\rmfamily]{Ligatures=TeX,Scale=1}
\fi
\usepackage{lmodern}
\ifPDFTeX\else
\fi
\IfFileExists{upquote.sty}{\usepackage{upquote}}{}
\IfFileExists{microtype.sty}{
  \usepackage[]{microtype}
  \UseMicrotypeSet[protrusion]{basicmath} 
}{}
\makeatletter
\@ifundefined{KOMAClassName}{
  \IfFileExists{parskip.sty}{%
    \usepackage{parskip}
  }{
    \setlength{\parindent}{0pt}
    \setlength{\parskip}{6pt plus 2pt minus 1pt}}
}{
  \KOMAoptions{parskip=half}}
\makeatother
\usepackage{xcolor}
\usepackage{longtable,booktabs,array}
\usepackage{multirow}
\usepackage{calc} 
\usepackage{etoolbox}
\makeatletter
\patchcmd\longtable{\par}{\if@noskipsec\mbox{}\fi\par}{}{}
\makeatother
\IfFileExists{footnotehyper.sty}{\usepackage{footnotehyper}}{\usepackage{footnote}}
\makesavenoteenv{longtable}
\usepackage{graphicx}
\makeatletter
\def\maxwidth{\ifdim\Gin@nat@width>\linewidth\linewidth\else\Gin@nat@width\fi}
\def\maxheight{\ifdim\Gin@nat@height>\textheight\textheight\else\Gin@nat@height\fi}
\makeatother
\setkeys{Gin}{width=\maxwidth,height=\maxheight,keepaspectratio}
\makeatletter
\def\fps@figure{htbp}
\makeatother
\ifLuaTeX
  \usepackage{selnolig}  
\fi
\IfFileExists{bookmark.sty}{\usepackage{bookmark}}{\usepackage{hyperref}}
\IfFileExists{xurl.sty}{\usepackage{xurl}}{} 
\hypersetup{
  hidelinks,
  pdfcreator={LaTeX via pandoc}}

\author{}
\date{}

\usepackage[a4paper,margin=1in]{geometry}
\usepackage{newtxtext,newtxmath}
\usepackage[fontsize=10.2pt]{fontsize}
\usepackage{microtype}
\usepackage{array}
\usepackage{colortbl}
\usepackage{hhline}
\usepackage{xurl}
\definecolor{tblhdr}{RGB}{197,217,241}
\usepackage{enumitem}
\setlist{topsep=2pt,itemsep=2pt,parsep=0pt,partopsep=0pt,leftmargin=1.5em,labelsep=0.4em,itemindent=0pt}
\usepackage{newunicodechar}
\newunicodechar{×}{\ensuremath{\times}}
\newunicodechar{·}{\ensuremath{\cdot}}
\newunicodechar{∈}{\ensuremath{\in}}
\newunicodechar{η}{\ensuremath{\eta}}
\newunicodechar{τ}{\ensuremath{\tau}}
\newunicodechar{ρ}{\ensuremath{\rho}}
\newunicodechar{Δ}{\ensuremath{\Delta}}
\newunicodechar{∧}{\ensuremath{\wedge}}
\newunicodechar{−}{\ensuremath{-}}
\newunicodechar{→}{\ensuremath{\rightarrow}}
\newunicodechar{↔}{\ensuremath{\leftrightarrow}}
\newunicodechar{⌈}{\ensuremath{\lceil}}
\newunicodechar{⌉}{\ensuremath{\rceil}}
\newunicodechar{≈}{\ensuremath{\approx}}
\newunicodechar{≥}{\ensuremath{\geq}}
\newunicodechar{≤}{\ensuremath{\leq}}
\newunicodechar{─}{\textendash}
\newunicodechar{└}{\ensuremath{\llcorner}}
\newunicodechar{₁}{\textsubscript{1}}
\newunicodechar{₂}{\textsubscript{2}}
\newunicodechar{¹}{\textsuperscript{1}}
\newunicodechar{²}{\textsuperscript{2}}
\newunicodechar{³}{\textsuperscript{3}}
\newunicodechar{⁴}{\textsuperscript{4}}
\newunicodechar{⁸}{\textsuperscript{8}}
\newunicodechar{⁺}{\textsuperscript{+}}
\usepackage{titlesec}
\titlespacing*{\section}{0pt}{9pt plus 2pt minus 1pt}{3pt}
\titlespacing*{\subsection}{0pt}{5pt plus 2pt minus 1pt}{2pt}
\titlespacing*{\subsubsection}{0pt}{4pt plus 1pt minus 1pt}{2pt}
\title{From Roofline to Ruggedness: Decomposing and Smoothing the GEMM Performance Landscape}
\author{\normalsize Aditya Chatterjee\\[2pt] \normalsize Intel Corporation\\[2pt] \normalsize \texttt{aditya.chatterjee@intel.com}}

\begin{document}
\maketitle

\begin{center}\rule{0.92\linewidth}{0.4pt}\end{center}

\begin{abstract}
\setlength{\parskip}{0.8\baselineskip plus 2pt}

Adjacent GEMM problems that differ by a single 128-element step in N can show 30\% different throughput on the same GPU. This pervasive \emph{performance ruggedness -} invisible to roofline analysis and peak-FLOPs intuition, yet dominant for every non-peak workload - is the subject of this paper.

We propose \textbf{performance ruggedness analysis} as an analytical framework complementary to roofline: rather than summarizing GPU performance with a scalar bound, treat the full multidimensional performance surface as the object of study, decomposes its texture into mechanism-attributable components, and separates software-removable losses from hardware-bound ones. The framing is directly analogous to deep-learning loss landscapes - a continuous quantity (the idealized time 2MNK / compute\_throughput\_peak) made rugged by interaction with discrete hardware substrates (tiles, sub-groups, cache lines, DRAM channels).

We apply the framework to BF16 NN (no-transpose) GEMM on Intel Battlemage (Arc B580, sycl-tla) via a 32,768-configuration sweep over (M, N, K) ∈ \{128, \ldots, 4096\}³. Peak throughput is 110.8 TFLOPs at the non-square shape M = 3840, N = 2048, K = 4096; to quantify texture we \textbf{introduce the concept of roughness}, the mean absolute step-to-step change in throughput, which starts at 16.8 TFLOPs per 128-step against an ideal of 2.0. A two-stage software stack - best-of-six dynamic tile selection and a \textbf{novel dynamic-programming padding-and-splitting optimizer} (precomputed once, applied at runtime as an O(1) lookup) - \textbf{reduces roughness by 70\% and raises mean throughput by 30\%}. Cross-tile experiments show the residual sawtooth period scales exactly with the software tile size, ruling out cache set-conflicts and attributing the remainder to four hardware-bound sources (wave-fill ramp and quantization, per-kernel base overhead, DPAS atom geometry, and GDDR6 channel-hash interactions). Finally, we \textbf{derive the optimal achievable landscape from first principles} \textbf{-} datasheet integers alone with no kernel run or simulator and turn it into an \textbf{optimality scale (Kernel Optimality Levels)} that grades any kernel by \textbf{how much of that landscape it attains and how close its roughness lies to the hardware floor}; on it, the production kernel and our optimized stack rate L0 and L2 respectively despite both reporting \textasciitilde95\% of theoretical hardware-bound peak.

\end{abstract}

\noindent\textbf{Keywords:} GEMM, Performance Modelling, GPU performance optimization, Intel Battlemage, GPU Microarchitecture, Dynamic Programming Optimization.

\section{Introduction}

\includegraphics[width=\linewidth]{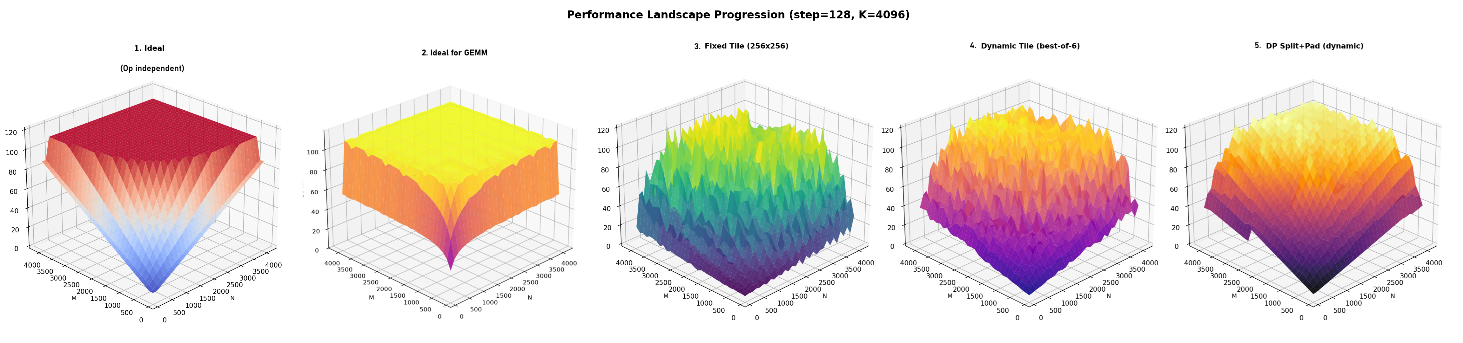}

Figure 1. Performance ruggedness on Intel BMG GEMM, before and after our optimization stack (step-128 resolution at K = 4 096). Left to right: \textbf{ideal-compute baseline independent of operation} (smooth pyramid; mean 97.2 TFLOPs, roughness 2.0 TFLOPs/step); \textbf{ideal-compute baseline for GEMM} (from first principles); \textbf{raw measured surface with the default 256x256 tile} (mean 58.7, roughness 16.8, an 8.2x ruggedness ratio); after \textbf{dynamic best-of-six tile selection} (mean 68.3, roughness 10.4); after our \textbf{novel dynamic-programming padding-and-splitting optimizer} applied on top of dynamic tile (mean 76.3, roughness 5.0 - a 70 \% reduction in landscape roughness and 30\% increase in mean TFLOPs). The ``ideal-roughness floor'' of 2.0 TFLOPs/step is itself a hardware consequence: it is the unavoidable monotone-ramp slope from low achievable throughput at small problems (limited by partial-wave fill on the 20 silicon-fixed Xe-cores) to peak at saturation. The remaining gap above this hardware-bound ramp is attributed in §8.5 to four additional hardware-bound sources (wave-quantization oscillation, DPAS atom asymmetry, per-kernel overhead variation and channel-hash interactions). All 5 TFLOPs / step of residual roughness is hardware-bound.

\subsection{The Roofline Hides a Surface}
Every modern GPU has a theoretical peak compute FLOPs value, but the achieved compute performance for any specific workload depends on its arithmetic intensity (FLOPs per byte of memory accessed) and peak FLOPs is limited by throttling. Roofline analysis {[}1{]} formalizes this dependence with two scalars - peak compute and peak memory bandwidth - bounding the maximum achievable performance for any given arithmetic intensity. In practice, even this roofline ceiling is rarely reached: real workloads fall short due to overhead beyond what roofline captures (kernel launch, sub-group barriers, partial-tile waste, register spill, imperfect compute-memory pipelining). The gap between the roofline ceiling and measured throughput is what this paper addresses.

On Intel Battlemage GPU (Arc B580), the achieved BF16 NN GEMM throughput varies by 30\% or more between adjacent (M, N, K) configurations that differ by a single 128-element step. The achieved performance is neither a scalar nor a roofline-clipped curve: it is a \textbf{multidimensional, rugged surface} and that ruggedness (not the peak) is what dominates the performance of any non-peak shape. This pattern is universal to all GPUs and CPUs.

We propose \textbf{performance ruggedness analysis} as a complementary analytical framework alongside the roofline model. Where the roofline says, \emph{``you cannot exceed this bound''}, ruggedness analysis says, \emph{``across the input space, here is where you fall short and why.''} The shift from a scalar bound to the full surface as the object of study is the central methodological contribution of this paper.

\subsection{Performance Landscape vs Loss Landscape}
The GEMM performance surface is structurally analogous to the loss landscapes of deep learning training optimization {[}2{]}: both are continuous domain quantities (loss in weight space; the idealized GEMM time 2MNK / peak\_theoretical\_FLOPs) made rugged by their interaction with fixed-size discrete units of the underlying machinery - in deep learning, these are mini-batches and floating-point arithmetic; in GEMM, these are tile sizes, sub-group widths, cache lines, and DRAM channels. We adopt this analogy as organizing structure throughout the paper.

\subsection{What We Did and What We Found}
We treat the BMG GEMM performance landscape as the primary object of study and answer four questions in sequence.

(1) \textbf{What does the landscape look like?} A 32,768 configuration sweep over (M, N, K) ∈ \{128, 256, \ldots, 4096\}\textsuperscript{3} reveals a mesa-shaped surface with peak 110.8 TFLOPs at non-square shape M=3840, N=2048, K=4096 with the default 256x256-tile kernel; the cross-tile peak rises to 112.1 TFLOPs at M=3072, N=2560, K=3840. Only 1.4 \% of configurations exceed 100 TFLOPs.

(2) \textbf{What creates the texture?} A four-surface decomposition (compute / memory / GEMM / residual overhead), built from paired microbenchmarks on the same grid, attributes each visible feature to a specific subsystem. Two previously undescribed measurement artifacts emerge: co-allocation memory interference (up to 50 \% slowdown when reading one of several simultaneously allocated buffers) and a 43 \% TLB/L3 temporal warmup drift across sequential measurement blocks. Both are eliminated by a randomized-order sweep which simultaneously reveals that within-configuration variation across 50 independent runs is 0.04 \% to 0.10 \% - the landscape is deterministic, not stochastic.

(3) \textbf{How much can software remove?} A two-stage stack: best-of-six dynamic tile selection (a standard technique) plus a novel dynamic-programming based padding-and-splitting optimizer \textbf{reduces landscape roughness by 70\%} (16.8 to 5.0 TFLOPs per 128-step) and raises mean TFLOPs by 30 \% (Figure 1, right).

(4) \textbf{What is left, and why?} Cross-tile fine-N sweeps at step-32 across three tile sizes (64x64, 128x128, 256x256) show that the sawtooth period equals the software tile size at every scale, with valleys uniformly at N mod tile\_size = 24; this establishes the periodic structure of the residual is partial-tile waste at progressively finer scales rather than any cache-conflict effect.

The remaining ruggedness - the entire 5 TFLOPs / step residual after our software stack - is attributable to four silicon-fixed hardware sources: the wave-fill ramp from partial-wave underutilization at small problems (the silicon-fixed C = 20 Xe-core count sets an unavoidable \textasciitilde2 TFLOPs / step monotone climb from small-problem throughput to peak) along with wave-quantization oscillation on top of the ramp; configuration-dependent variation in the per-kernel \textasciitilde32\% base overhead (rooted in the fixed 256-GRF register file, sub-group barrier latency, and GDDR6 latency-bound load semantics); the 8 x 16 x 16 DPAS atom geometry's M:N asymmetry; and GDDR6 channel-hash interaction with (M, N) memory layouts.

\subsection{Generality of the framework}
The performance-ruggedness framing, the four-surface decomposition, the randomized-order sweep methodology and the dynamic-programming padding-and-splitting optimizer are all platform-independent constructions: each operates on a measured timing table T0{[}M{]}{[}N{]}{[}K{]} and assumes no Intel-specific hardware feature. The four mechanism classes they expose - partial-tile waste at tile boundaries, scheduling tail effects on a fixed unit count, memory-channel contention and TLB / cache warmup drift - are intrinsic to any tile-based dense matrix kernel running on a parallel processor with multi-channel memory and virtual addressing. This includes other GPU architectures (NVIDIA Hopper / Blackwell, AMD MI300), CPU GEMM kernels using SIMD / AVX-512 register tiles with multi-core scheduling and NUMA / multi-channel DRAM and tile-based AI accelerators.

The present paper instantiates the framework on Intel Battlemage GPU (Arc B580); the specific quantitative magnitudes - initial roughness 16.8 TFLOPs / step, residual 5.0, the 32\% per-kernel base overhead, the 50 \% co-allocation interference, the 43\% temporal drift - are BMG-specific and will shift on other platforms. The qualitative landscape signatures (mesa shape, tile-period sawtooth, configuration-dependent overhead variation) are universal; cross-platform validation is the natural extension and is left to future work.

\section{Background: Intel Battlemage, GEMM Tile Hierarchy and Setup}

We measure on Intel Battlemage (Arc B580), a GPU implementing the Intel Xe2 microarchitecture. Hardware parameters (Table 1) are from Intel architecture disclosures.

\subsection{BMG G21 (Arc B580)}
\begin{longtable}[]{!{\color{black}\vrule}
  >{\raggedright\arraybackslash}p{(\columnwidth - 2\tabcolsep) * \real{0.5019}}!{\color{black}\vrule}
  >{\raggedright\arraybackslash}p{(\columnwidth - 2\tabcolsep) * \real{0.4981}}!{\color{black}\vrule}}
\hhline{|-|-|}
\cellcolor{tblhdr}\begin{minipage}[b]{\linewidth}\raggedright
\textbf{Property}
\end{minipage} & \cellcolor{tblhdr}\begin{minipage}[b]{\linewidth}\raggedright
\textbf{Value}
\end{minipage} \\
\hline
\endhead

\endlastfoot
\textbf{Xe-cores} & 20 \\ \hline
\textbf{Vector engines per Xe-core} & 8 \\ \hline
\textbf{Total hardware threads} & 160 \\ \hline
\textbf{SIMD width (sub-group)} & 16 lanes \\ \hline
\textbf{DPAS atom} & 8 x 16 x 16 \\ \hline
\textbf{Peak BF16 throughput (theoretical)} & 116.5 TFLOP/s \\ \hline
\textbf{GRF register file per thread} & 256 × 32 B = 8 KB \\ \hline
\textbf{L1 / shared local memory} & 64 KB per Xe-core \\ \hline
\textbf{L3 cache (shared)} & 12 MB \\ \hline
\textbf{Memory technology} & GDDR6, 12 GB, 192-bit, \textasciitilde456 GB/s peak \\ \hline
\textbf{Memory channels} & 6 \\ \hline
\end{longtable}

\begin{quote}
Table 1. Intel Battlemage G21 (Arc B580) parameters.
\end{quote}

\subsection{GEMM Tile Hierarchy on BMG}
The sycl-tla port {[}4{]} of CUTLASS {[}5{]} organizes a GEMM \textbf{C = A · B} (with A: M × K, B: K × N) as five nested tiles:

\begin{quote}
{\ttfamily\setlength{\parindent}{0pt}%
Output matrix C\quad(M\,$\times$\,N)\\
\hspace*{2em}$\hookrightarrow$~Work-group tile\quad 256$\times$256$\times$32\quad(default)\\
\hspace*{4em}$\hookrightarrow$~Sub-group tile\quad 32$\times$64\quad(8 SGs along M, 4 along N)\\
\hspace*{6em}$\hookrightarrow$~DPAS atom\quad 8$\times$16$\times$16\quad(M$\times$N$\times$K-inner)\\
\hspace*{8em}$\hookrightarrow$~SIMD lane\quad 1$\times$16\quad(one sub-group)%
}

Each workgroup writes a 256x256 chunk of output C and contains 32 subgroups arranged in an 8x4 grid; each subgroup produces a 32x64 sub-tile. Within a sub-group, the 32x64 output is computed as 4x4 = 16 DPAS calls per K=16 chunk. Operand B is loaded into GRFs in VNNI layout by a single 2D block load instruction that fetches a 64-column x 32-K-row slab pre-permuted for direct DPAS consumption.
\end{quote}

\subsection{The K-Iteration Mainloop}
The kernel iterates over K/32 = 128 K-blocks (for K=4096). Each iteration overlaps a 2-stage prefetch ahead of the consuming DPAS chain:

1. \textbf{PREFETCH} non-blocking hints for tiles 2 K-iterations ahead.

2. \textbf{LOAD} A{[}256, 32{]} and B{[}32, 256{]} from DRAM via L3/L1 into GRF.

3. \textbf{DPAS} systolic compute on the loaded tiles, producing output in registers.

4. \textbf{BARRIER} subgroup synchronization at the end of the iteration.

5. After all K iterations, \textbf{STORE} the 256x256 output through L1/L3/GDDR6.

The stages can overlap and not serialize. The 2-stage prefetch hides load latency only when:

\begin{itemize}
\item
  Sufficient K-iterations exist to fill the pipeline (small K leaves the pipeline empty)
\item
  GRF budget can hold two in-flight tile loads plus the accumulator without spilling.
\end{itemize}

The kernel handles non-tile-aligned dimensions via ceil\_div in the tile scheduler. For M=300 with tile\_M=256, two workgroups are launched in M; the second processes 44 valid rows out of 256 and the epilogue uses residue coordinates to discard out-of-bounds output.

We use a uniform 128-element step in each dimension for two reasons:

\begin{itemize}
\item
  It bounds the total measurement count to 32,768 - a tractable wall-clock budget while still preserving multiples of all common tile widths (128, 256, 512, 1024).
\item
  We complement the full 3D sweep with fine-grained 1D sweeps at step = 32 along the N axis (Sections 6.3 and 8.3), which reveal that the same partial-tile-waste mechanism produces sawtooth periods that scale exactly with the chosen tile size - there are no qualitatively new mechanisms at finer resolution, only smaller-amplitude versions of the same mechanism.
\end{itemize}

After applying the DP optimizer at the fine grid, residual roughness drops to \textasciitilde1.31 TFLOPs/step (Table 14), approximately 22x the within-configuration measurement noise. We therefore use step = 128 for the 32,768-configuration sweep (which covers the full (M, N, K) cube) and step = 32 for targeted mechanism-refinement experiments.

All measurements were taken on a single workstation with an Intel Arc B580 (Battlemage G21) and an Intel Core i5-13400 host CPU running Ubuntu 25.04 (kernel 6.15.0-061500-generic), with the discrete GPU on the i915 driver. The software stack is Intel oneAPI Base Toolkit 2025.1 (icpx 2025.1.1), Level Zero loader 1.28.0 with the Intel Compute Runtime (libze-intel-gpu1 26.05.37020.3, intel-igc-core-2 2.28.4), and sycl-tla on branch ``main'' (commit be58860e) built with CUTLASS\_SYCL\_PROFILING\_ENABLED=ON. Compiler flags include ``-cl-intel-256-GRF-per-thread'' (full GRF budget) and ``IGC\_VectorAliasBBThreshold=100000000000'' (enables 2D-block-load alias analysis to complete). Hardware-counter access requires the kernel sysctls dev.i915.perf\_stream\_paranoid=0 and kernel.perf\_event\_paranoid=0.

We use the following definitions consistently.

\begin{itemize}
\item
  \textbf{Achieved throughput (TFLOPs)}: For a measurement of GEMM kernel time t at problem size (M, N, K), achieved throughput in tera-FLOPs per second is
\end{itemize}

\[\text{TFLOPs}(M,N,K) = \frac{2 \cdot M \cdot N \cdot K}{t \cdot 10^{12}}.\]

\begin{itemize}
\item
  \textbf{Mean TFLOPs}: Arithmetic mean of TFLOPs across all configurations within a stated subset (e.g., across all 32 768 configurations of the full sweep, or across the 35 N-values of the fine-N sweep at fixed M = K = 4096).
\item
  \textbf{Roughness}: We measure the local non-smoothness of the landscape along a one-dimensional axis (typically N at fixed M, K) as the mean absolute step-to-step difference in achieved TFLOPs:
\end{itemize}

\begin{quote}
\[\text{Roughness}\left( \text{\{}T_{1},T_{2},\ldots,T_{n}\text{\}} \right) = \frac{1}{n - 1}\sum_{i = 1}^{n - 1}\left| T_{i + 1} - T_{i} \right|\]

where T\textsubscript{i} is the TFLOPs measurement at the i\textsuperscript{th} value of the sweep axis. For a perfectly linearly varying surface, roughness is bounded below by the average per-step compute increase; for the step-128 sweep on the ideal-compute model 2MNK / P\textsubscript{peak}, this floor is 2.0 TFLOPs/step (slope of the ideal performance landscape for GEMM). We report roughness in TFLOPs / step.
\end{quote}

\begin{itemize}
\item
  \textbf{CV (coefficient of variation)}: Ratio of standard deviation \(\sigma\) to mean \(\mu\), expressed as a percent:\(\ \text{CV} = \sigma\text{/}\mu\  \times \ 100\%\). A CV of 0.05 \% indicates the measurement is essentially noise-free; a CV of 4 \% indicates moderate spread.
\item
  \textbf{Drift}: Systematic change in a measured value across an ordered sequence of measurements: example, ``43 \% drift'' means the value at the end of the sequence differs from the value at the start by 43\% in a consistent direction.
\item
  \textbf{Configs \textgreater{} X \%}: Number of (M, N, K) configurations (out of the stated total) where the effect under study exceeds the threshold X\%. For example, ``32\% of configs \textgreater{} 20 \% slowdown'' means 32\% of the 32,768 configurations have ≥ 20 \% slowdown.
\end{itemize}

Terminology: We use ``ruggedness'' to refer qualitatively to the phenomenon of non-smooth performance landscapes and ``roughness'' for the specific quantitative metric defined. Throughout the paper, ``the landscape is rugged'' denotes the phenomenon: ``roughness = 5 TFLOPs/step'' denotes the metric.

\section{GEMM Performance Landscape}

We construct the performance landscape by measuring BF16 GEMM throughput across a regular 3-dimensional grid of problem sizes. The axes are the matrix dimensions M, N and K; the value at each grid point is the throughput in TFLOP/s. The grid spans (M, N, K) ∈ \{128, 256, \ldots, 4096\} at uniform 128-element step in each dimension, yielding 32,768 measured configurations. Each configuration\textquotesingle s time is averaged over five repeated kernel launches with the first launch discarded as warmup, giving a clean signal whose stochastic component is below 0.1\%. Visualizing this 3D surface requires reduction to 2D so we collapse one axis and inspect the resulting (M, N) surface or its 1D slices.

Note: Five runs are sufficient because the within-configuration coefficient of variation is 0.04 -- 0.10\% (Table 12), giving a standard error of the mean below 0.05\% with 5 runs - well below the magnitudes of every effect we report. Larger run counts (50, used in Section 8.2) do not change conclusions.

\subsection{Mesa Shape and Three Performance Regimes}
The first observation is that the achieved throughput takes the form of a mesa with a sharply rising slope and a broad, textured plateau.

\includegraphics[width=2.58194in,height=2.24306in]{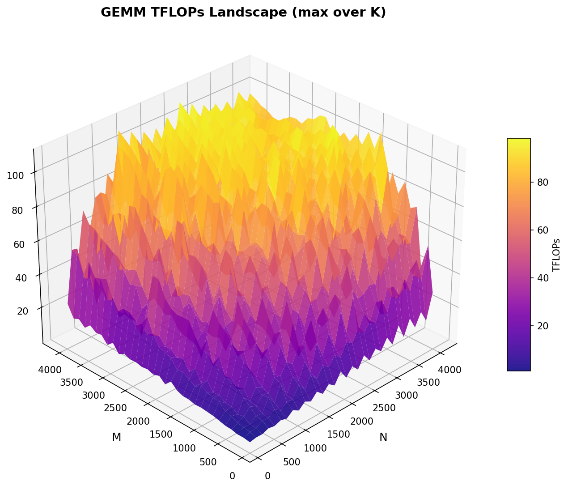} \includegraphics[width=3.12153in,height=2.03681in]{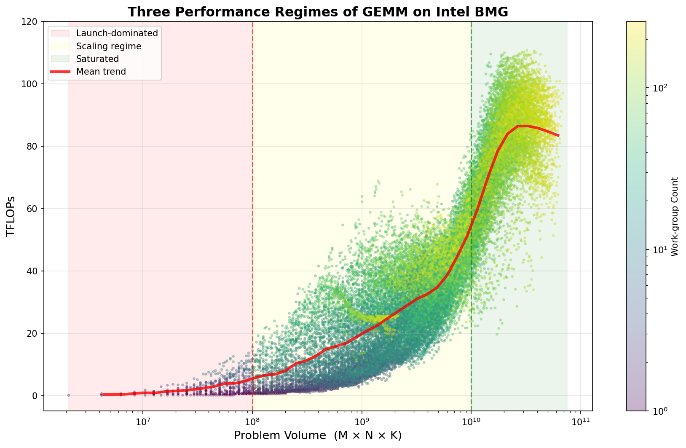}

Figure 2 (left): GEMM performance landscape, max over K. The peak in the entire 32,768-configuration sweep is 110.8 TFLOPs at M=3840, N=2048, K=4096. The plateau is broad but rugged: large regions of the surface sit between 80 and 110 TFLOPs separated by visible ridges and valleys.

Figure 2 (right): Achieved TFLOP/s as a function of problem volume. Each point is one of the 32,768 configurations. Figure 2 exposes a clean 3-regime separation:

\begin{longtable}[]{!{\color{black}\vrule}
  >{\raggedright\arraybackslash}p{(\columnwidth - 6\tabcolsep) * \real{0.2502}}!{\color{black}\vrule}
  >{\raggedright\arraybackslash}p{(\columnwidth - 6\tabcolsep) * \real{0.2460}}!{\color{black}\vrule}
  >{\raggedright\arraybackslash}p{(\columnwidth - 6\tabcolsep) * \real{0.2475}}!{\color{black}\vrule}
  >{\raggedright\arraybackslash}p{(\columnwidth - 6\tabcolsep) * \real{0.2562}}!{\color{black}\vrule}}
\hhline{|-|-|-|-|}
\cellcolor{tblhdr}\begin{minipage}[b]{\linewidth}\raggedright
\textbf{Regime}
\end{minipage} & \cellcolor{tblhdr}\begin{minipage}[b]{\linewidth}\raggedright
\textbf{Volume (MNK)}
\end{minipage} & \cellcolor{tblhdr}\begin{minipage}[b]{\linewidth}\raggedright
\textbf{TFLOPs (mean)}
\end{minipage} & \cellcolor{tblhdr}\begin{minipage}[b]{\linewidth}\raggedright
\textbf{\% of configurations}
\end{minipage} \\
\hline
\endhead

\endlastfoot
\textbf{Launch dominated} & \textless{} 10\textsuperscript{8} & \textless{} 5 & 17\% \\ \hline
\textbf{Scaling} & 10\textsuperscript{8} to 10\textsuperscript{10} & 5 to 80 & 65\% \\ \hline
\textbf{Saturated} & \textgreater{} 10\textsuperscript{10} & 80 to 110 & 18\% \\ \hline
\end{longtable}

Table 2. Three GEMM performance regimes.

In the launch-dominated regime, the kernel completes faster than its setup cost. In the scaling regime, the GPU is progressively filling: additional workgroup activates more of the 20 Xe-cores. In the saturated regime, all Xe-cores are busy, and additional workgroups extend the duration of execution. The transition into saturation occurs near (M/256)×(N/256) ≈ 60 workgroups roughly 3 full waves on the 20-core BMG.

Notably, the saturated plateau spans 80 to 110 TFLOPs (a \textasciitilde30 TFLOPs band) but only 1.4\% of all configurations exceed 100 TFLOPs. The plateau is broad in extent but sparse near its peak; understanding this gap is one motivation for the DP optimizer.

The two cutoff volumes (10\textsuperscript{8} and 10\textsuperscript{10}) are derived empirically from the achieved-throughput-vs-volume curve (Figure 2): below 10⁸, the curve is essentially flat at \textless{} 5 TFLOPs; between 10\textsuperscript{8} and 10\textsuperscript{10}, the curve rises sharply (the GPU is filling); above 10\textsuperscript{10}, the curve plateaus around 80-110 TFLOPs (saturated). These cutoffs are not theoretical but data-driven boundaries between qualitatively different regimes.

\subsection{Aspect-Ratio Sensitivity: Peak Is Not at the Largest Square}
A common assumption is that the largest square problem (M=N) maximizes throughput. The data refutes this on BMG. The square M = N = 4096 achieves 97.1 TFLOPs while the rectangular M = 3840, N = 2048, K = 4096 reaches 110.8 TFLOPs. To characterize this asymmetry across the full sweep, we plot mean TFLOP/s as a function of the aspect ratio M/N at fixed K.

\includegraphics[width=\linewidth]{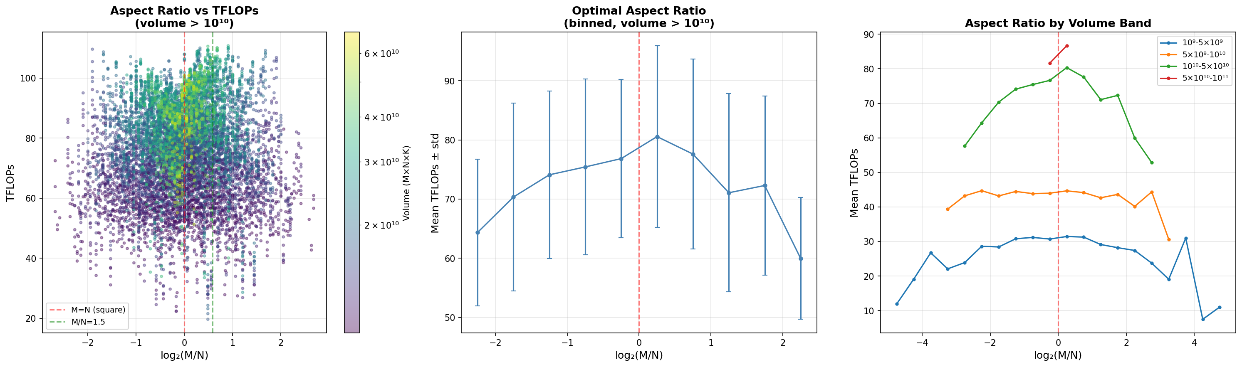}

Figure 3. Mean TFLOP/s as a function of M/N ratio at fixed K=4096. Aspect ratios near 1.0 (squares) underperform; the curve peaks in the range 1.0 to 1.5 (slightly rectangular, M \textgreater{} N) and degrades sharply for either highly elongated direction.

The optimal M:N ratio lies in the 1.0 to 1.5 range. Two factors contribute:

\begin{itemize}
\item
  The workgroup is intrinsically asymmetric (256 × 256 workgroups arranged 8 × 4 in sub-groups) so the kernel\textquotesingle s computational cost is not symmetric in M and N.
\item
  The GDDR6 {[}6{]} bandwidth is shared across reads of A (size M×K) and B (size K×N) and modest asymmetry between A\textquotesingle s and B\textquotesingle s footprints reduces channel contention compared to a perfectly balanced load.
\end{itemize}

Note: The single highest-throughput configuration (M=3840, N=2048) sits at M:N = 1.875 - it is not at the aggregate peak but is the highest single point because of partial-tile-waste alignment at this specific shape.

The implication is: for any GEMM workload where the user has freedom to choose the iteration ordering or to swap M and N (transposed dispatches), the rectangular shape with M slightly larger than N is preferable.

\subsection{Dimension Alignment: N matters 2.5X more than M}
The plateau\textquotesingle s texture is dominated by sharp performance jumps at specific values of M and N. To quantify these, we restrict observation to subsets of the grid where one dimension lies on a 256-boundary and compare the mean throughput against subsets where it does not.

\includegraphics[width=\linewidth]{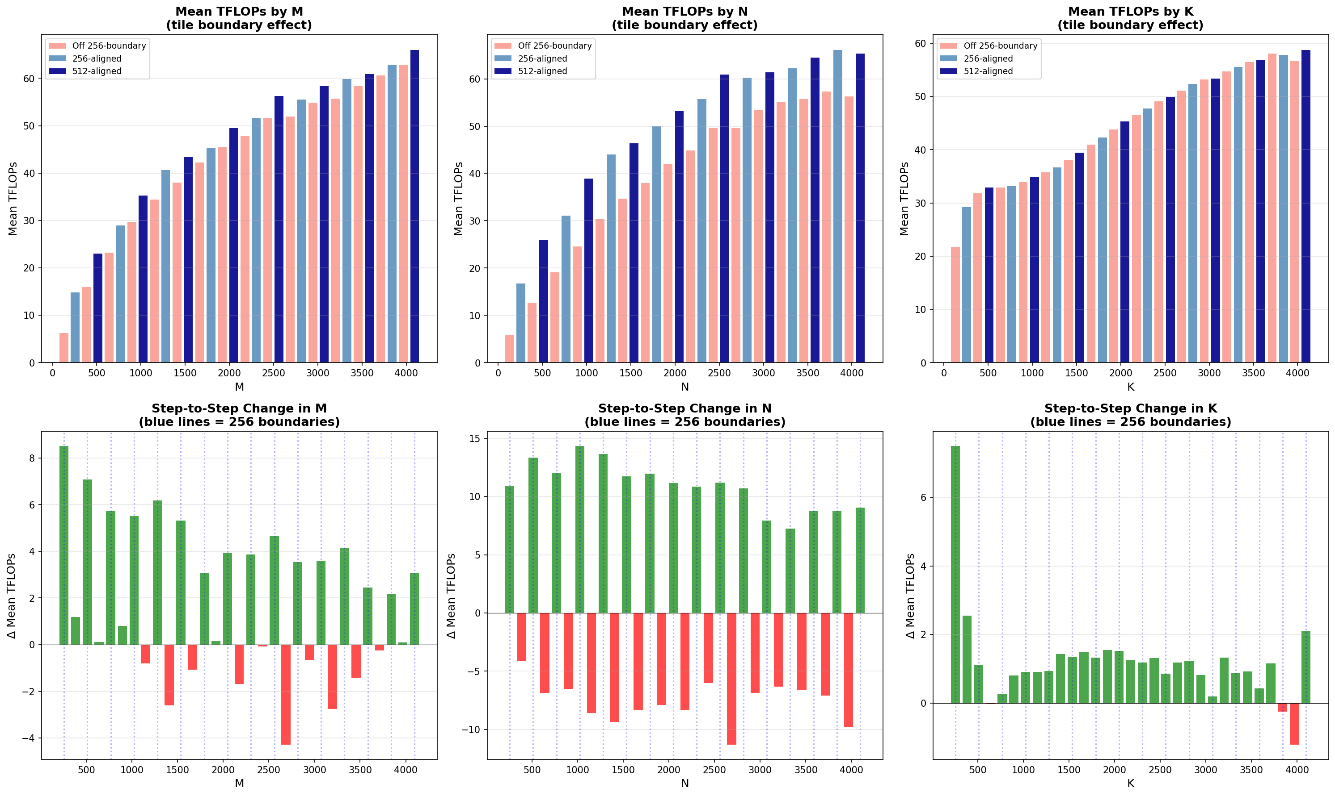}

Figure 4. Performance jumps when crossing 256-element boundaries. Top: aligning N to 256-multiple boosts mean TFLOPs by +27.6\% vs the immediately adjacent off-boundary value. Bottom: aligning M boosts mean TFLOPs by +10.7\%. The N-cliffs are roughly 2.5X the height of the M-cliffs, an asymmetry consistent across the entire sweep.

The asymmetry is rooted in the sub-group layout. Each workgroup contains 32 sub-groups arranged 8 × 4 so each sub-group handles a 32×64 output tile (32 rows of M x 64 columns of N). 3 consequences follow:

\begin{itemize}
\item
  Subgroup\textquotesingle s N-extent (64) is twice its M-extent (32) so non-64-aligned N wastes more compute per subgroup than non-32-aligned M.
\item
  DPAS atom produces 8 rows × 16 columns making N = 16 the atomic unit on the N axis. N-misalignment cascades through pipeline levels more than M-misalignment.
\item
  2D block load of operand B is laid out K-major so for small N, the load uses fewer cache lines per fetch reducing effective bandwidth.
\end{itemize}

Worked example: The single steepest jump in the entire landscape is at N = 1024 where mean throughput rises by 14.3 TFLOPs in a single 128-element step. At step-128 resolution, the relevant adjacent points are N ∈ \{768, 896, 1024, 1152\}. With work-group tile width 256, the kernel launches ⌈N/256⌉ workgroups in the N direction:

\begin{itemize}
\item
  N = 768: 3 WGs in N, all 100\% utilized, total 100\%.
\item
  N = 896: 4 WGs in N, last one only 50\% utilized, total 87.5\%.
\item
  N = 1024: 4 WGs in N, all 100\% utilized, total 100\%.
\item
  N = 1152: 5 WGs in N, last 50\%, total 90\%.
\end{itemize}

The transition 896 to 1024 keeps the work-group count constant (4 in N) and the wave structure unchanged; the only change is that the previously 50\%-utilized last workgroup becomes fully utilized. The arithmetic predicts a +12.5\% throughput gain (100/87.5 − 1), matching the measured +14.3 TFLOP/s within \textasciitilde10\%. The mechanism: partial-tile waste at work-group boundaries, recurs throughout the paper and forms the analytical core of the optimization stack.

\subsection{K Has Diminishing Returns}
The final dimension K behaves differently from M and N because the kernel processes K in inner-loop iterations rather than outer-loop workgroups, so increasing K does not change the workgroup count or the partial-tile waste pattern. Instead, K affects only the per-iteration amortization of fixed kernel-launch and pipeline-startup cost.

As a result, mean TFLOP/s increases sharply at small K and then plateaus from K = 128 to K = 1024, mean throughput rises by 38\%, but from K = 2048 to K = 4096, it rises by only 7\%. Beyond K ≈ 1024, the kernel is amortized, further K growth yields essentially constant throughput. This insensitivity makes K a natural splitting axis for our DP optimizer, since K-splits cost little throughput but expose more parallelism for compute-bound configurations.

\section{Decomposition: Where Does Texture Come From?}

The GEMM performance landscape is rich but inscrutable: which subsystem creates each visible feature? We attribute the texture by measuring four surfaces over the same (M, N, K) grid:

\begin{itemize}
\item
  \textbf{Compute surface}: \(t_{compute}\) = 2·M·N·K / 116.5 ·10¹² (roofline-style ideal: the time the workload would take if every issued FLOP produced useful output at peak BF16 throughput; depends on problem dimensions and the hardware compute peak). This does not depend on tile shape, kernel implementation or memory subsystem as it considers useful FLOPs only (not issued FLOPs).
\item
  \textbf{Memory surface}: \(t_{memory}\) = performing the same A-load + B-load + D-store traffic the GEMM kernel would but with no DPAS work. This captures real cache effects, allocator behaviour and channel utilization.
\item
  \textbf{GEMM surface}: \(t_{GEMM}\) = measured GEMM time. This includes all of compute, memory, prefetch, barriers, kernel launch and partial-tile waste.
\item
  \textbf{Overhead surface}: \(t_{overhead} = t_{GEMM} - max\left( t_{compute},t_{memory} \right)\). This captures imperfect compute-memory pipelining, sub-group barrier latency, register spill, partial-tile waste, and scheduling overhead.
\end{itemize}

Partial-tile waste the FLOPs the hardware issues but the kernel discards because M, N or K are not tile-multiples and is therefore not absorbed into compute; it appears in the overhead surface. This separation is intentional: it keeps the compute baseline tile-independent so that the same decomposition is comparable across the dynamic six tile choices and across the pre/post-DP split/pad comparison.

The memory microbenchmark uses the same 2D block-load instruction sequence the GEMM kernel would use, but with the DPAS computation replaced by a no-op. This ensures the access-pattern stride, cache-line utilization, and memory-controller request distribution match GEMM\textquotesingle s actual memory traffic. Differences between \(t_{memory}\) and the memory-only portion of \(t_{GEMM}\) therefore reflect

\begin{itemize}
\item
  imperfect compute-memory pipelining in the GEMM kernel, captured in \(t_{overhead}\)
\item
  microbenchmark-vs-GEMM access-pattern differences, which we minimize by using the same load instructions.
\end{itemize}

\subsection{The Four Component Surfaces}
\includegraphics[width=\linewidth]{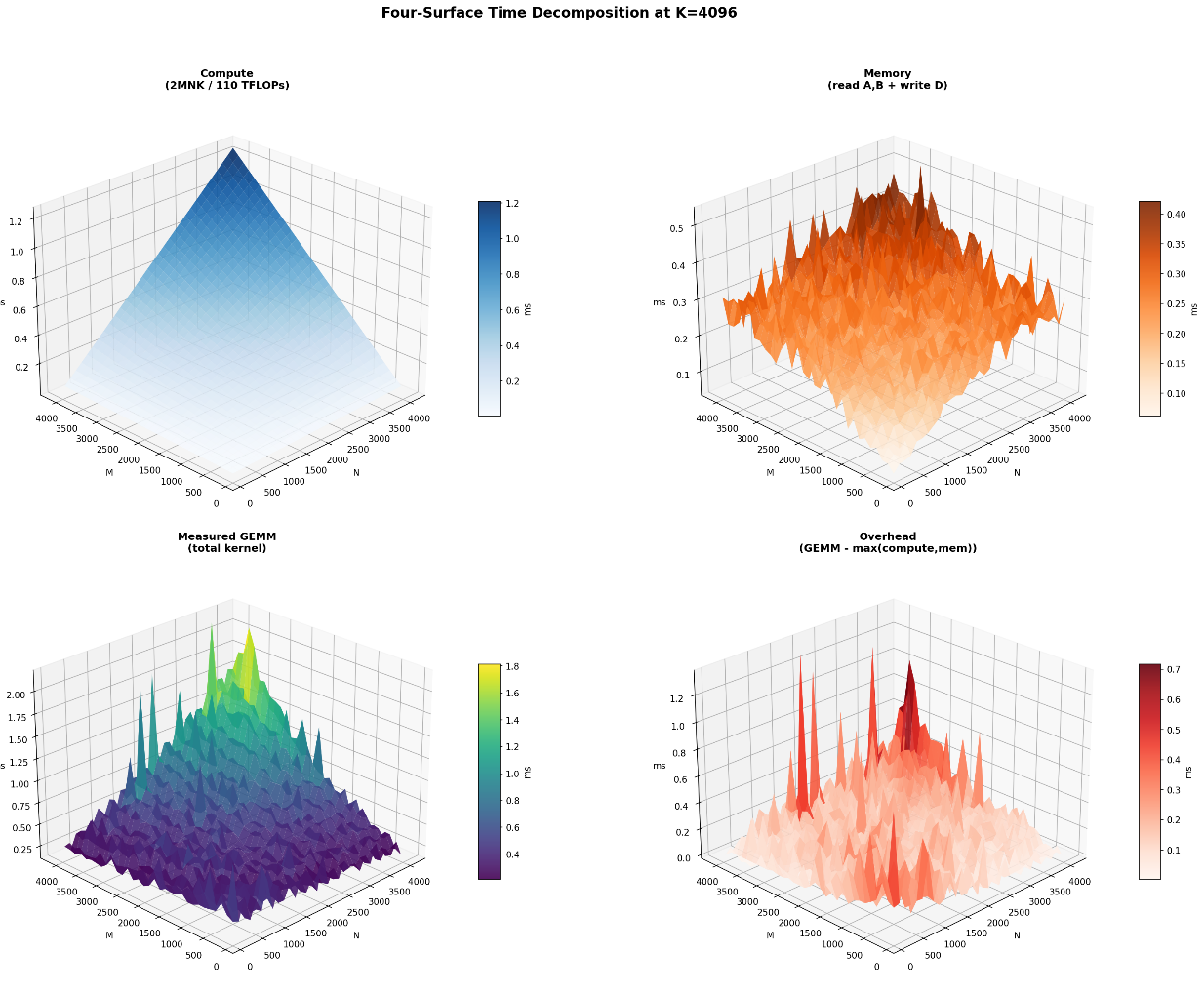}

Figure 5. Decomposition at K = 4096. Top-left: pure compute time - a perfectly smooth pyramid (no texture, by construction). Top-right: measured memory time - visibly jagged, with checkerboard texture from cache effects. Bottom-left: measured GEMM time - combines memory texture with isolated overhead spikes. Bottom-right: residual overhead t\_GEMM=max(t\_compute, t\_memory) - sparse hot spots at specific (M, N) plus a near-constant base level.

The surfaces tell us immediately: the TFLOPs surface inherits all its ruggedness from memory and overhead. Memory contributes the bulk of the structured texture (checkerboard, step patterns at cache boundaries); overhead contributes spike-like outliers at specific shapes plus a roughly constant base.

\subsection{The Constant 32\% Overhead Floor}
\includegraphics[width=\linewidth]{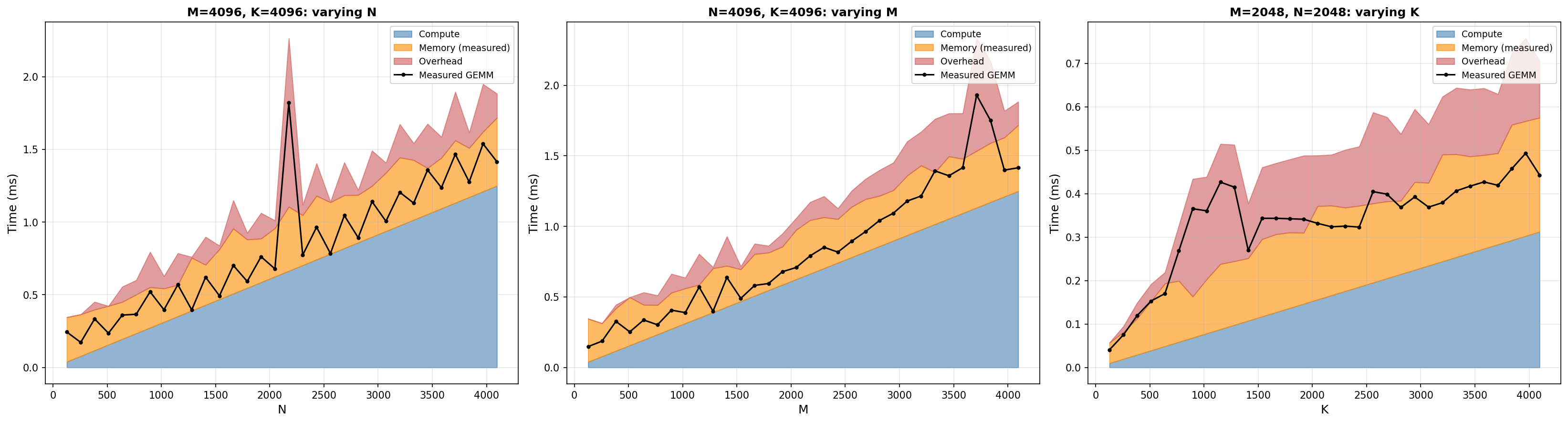}

Figure 6. Stacked time decomposition along varying N (M, K = 4096).

Memory dominates at small N; compute catches up at large N; the overhead bar (red) hovers near 32\% of total time across the full range. This constancy is the basis of the 32\% per-kernel base-overhead claim referenced throughout the paper. The 32\% is a composite of pipeline startup/drain, sub-group barrier latency, register-pressure-induced spill, and partial-tile waste.

\subsection{Roofline and Bottleneck Classification}
\begin{longtable}[]{!{\color{black}\vrule}
  >{\raggedright\arraybackslash}p{(\columnwidth - 4\tabcolsep) * \real{0.3899}}!{\color{black}\vrule}
  >{\raggedright\arraybackslash}p{(\columnwidth - 4\tabcolsep) * \real{0.2777}}!{\color{black}\vrule}
  >{\raggedright\arraybackslash}p{(\columnwidth - 4\tabcolsep) * \real{0.3324}}!{\color{black}\vrule}}
\hhline{|-|-|-|}
\cellcolor{tblhdr}\begin{minipage}[b]{\linewidth}\raggedright
\textbf{Bandwidth used in roofline}
\end{minipage} & \cellcolor{tblhdr}\begin{minipage}[b]{\linewidth}\raggedright
\textbf{Compute bound}
\end{minipage} & \cellcolor{tblhdr}\begin{minipage}[b]{\linewidth}\raggedright
\textbf{Memory bound}
\end{minipage} \\
\hline
\endhead

\endlastfoot
\textbf{Theoretical (456 GB/s, GDDR6 spec)} & 55\% & 45\% \\ \hline
\textbf{Measured (\textasciitilde270 GB/s sequential)} & 32\% & 68\% \\ \hline
\end{longtable}

Table 3. Bottleneck classification depends critically on which bandwidth one uses.

The classification compares each configuration\textquotesingle s roofline-ideal compute time to its measured memory time - a configuration is ``compute-bound'' if even a hypothetical zero-waste, perfectly pipelined kernel would be compute-limited. Achieved bandwidth is far below the GDDR6 spec as observed on multi-channel parallel hardware {[}7{]}.

\section{Memory Subsystem Analysis}

Detailed GPU memory characterization {[}8{]} is a long-established research tradition. We extend it with a sweep-scale methodology that uncovers two previously undescribed measurement artifacts, then characterize them in detail. The memory surface is the most jagged of the four. We characterize it in detail and in the process documenting two previously undescribed measurement artifacts and the randomized-sweep methodology that resolves them. The bandwidth microbenchmark is run in two variants:

\begin{itemize}
\item
  with each buffer co-allocated (mimicking the GEMM kernel\textquotesingle s address space)
\item
  with each buffer isolated (one malloc\_device per measurement).
\end{itemize}

The discrepancy between the two variants is the source of the first finding.

\subsection{Load vs Store Asymmetry}
Across the (M, N, K) grid for identical FP32 buffers (Figure 7), writes achieve 324 GB/s mean (max 662 GB/s) vs reads at 275 GB/s mean (max 557 GB/s) - a 1.18x write/read ratio consistent across all data sizes. The asymmetry is the expected signature of write-combining: GPU memory controllers coalesce small stores into burst writes while reads are latency-bound consuming EU stalls until the data arrives.

\includegraphics[width=4.98836in,height=4.31079in]{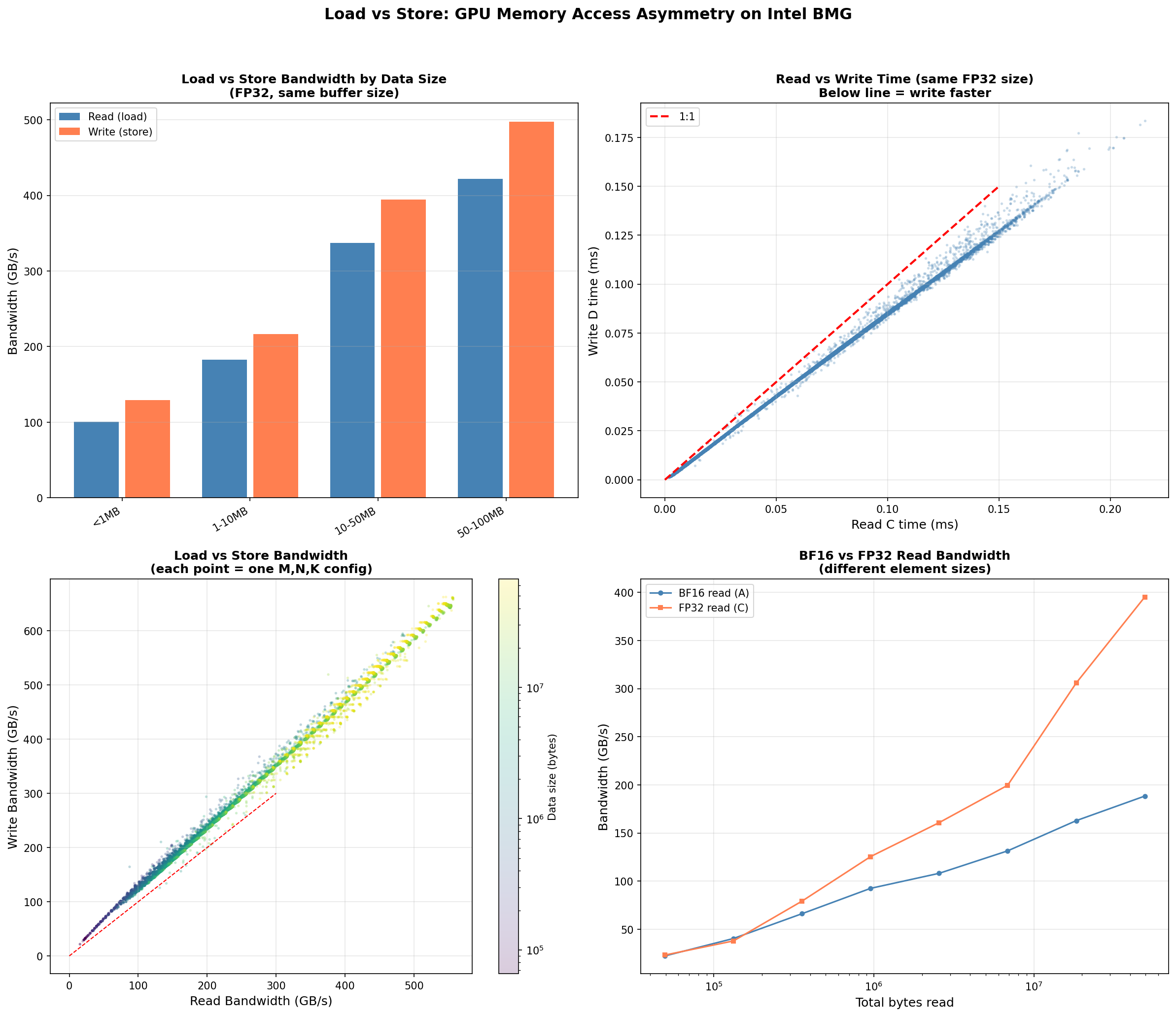}

Figure 7. Read vs write bandwidth distributions for identical FP32 buffers, demonstrating the consistent 18\% write-combining advantage.

\subsection{Co-Allocation Memory Interference}
Co-allocation interference is a property of any multi-channel GPU memory subsystem: when several buffers share an address space served by the same memory controller, requests can compete for individual channels. The phenomenon is universal across modern GPU architectures (HBM, GDDR), but the \textbf{magnitude and shape of the interference is a measurement that has not, to our knowledge, been characterized at sweep scale}. We do so on BMG GDDR6.

When the same single buffer A is read in two contexts - alone vs simultaneously allocated alongside B, C, D - its read time differs systematically.

\includegraphics[width=\linewidth]{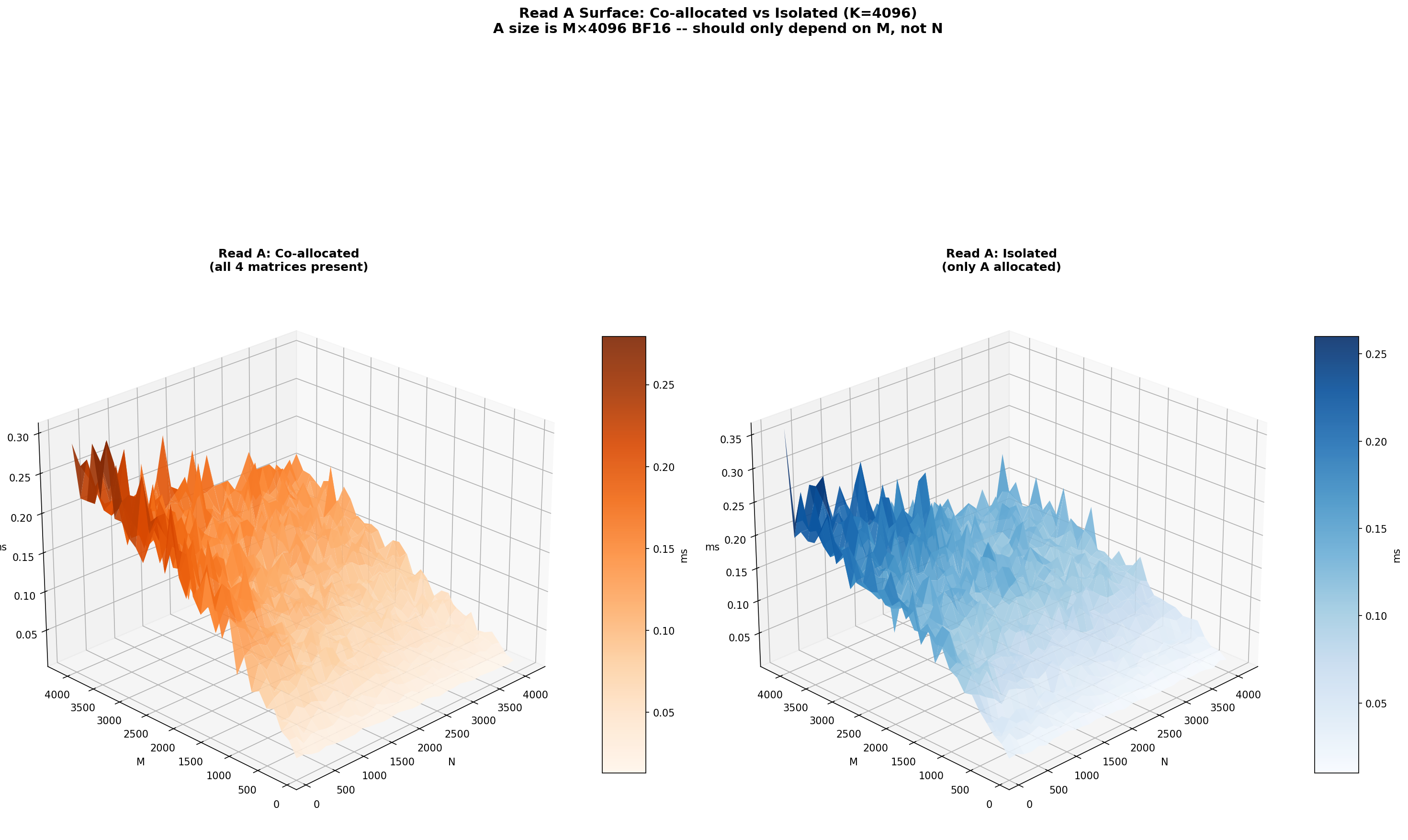}

Figure 8. Read-A time as a function of (M, N). Left: A is co-allocated with B, C, D - read-A time depends on N (which controls B, C, D sizes). Right: A is isolated - read-A time depends only on M (which controls A\textquotesingle s own size). The surfaces are qualitatively different.

\begin{longtable}[]{!{\color{black}\vrule}
  >{\raggedright\arraybackslash}p{(\columnwidth - 6\tabcolsep) * \real{0.2477}}!{\color{black}\vrule}
  >{\raggedright\arraybackslash}p{(\columnwidth - 6\tabcolsep) * \real{0.2509}}!{\color{black}\vrule}
  >{\raggedright\arraybackslash}p{(\columnwidth - 6\tabcolsep) * \real{0.2507}}!{\color{black}\vrule}
  >{\raggedright\arraybackslash}p{(\columnwidth - 6\tabcolsep) * \real{0.2507}}!{\color{black}\vrule}}
\hhline{|-|-|-|-|}
\cellcolor{tblhdr}\begin{minipage}[b]{\linewidth}\raggedright
\textbf{Buffer}
\end{minipage} & \cellcolor{tblhdr}\begin{minipage}[b]{\linewidth}\raggedright
\textbf{Mean slowdown}
\end{minipage} & \cellcolor{tblhdr}\begin{minipage}[b]{\linewidth}\raggedright
\textbf{Configs \textgreater{} 20\%}
\end{minipage} & \cellcolor{tblhdr}\begin{minipage}[b]{\linewidth}\raggedright
\textbf{Configs \textgreater{} 50\%}
\end{minipage} \\
\hline
\endhead

\endlastfoot
Read A & 1.12X & 32\% & 10\% \\ \hline
Read B & 1.11X & 31\% & 10\% \\ \hline
Write D & 1.11X & 30\% & 9\% \\ \hline
\end{longtable}

Table 4. Co-allocation interference impact across reads and writes.

Co-allocated buffers compete for memory-controller channels. At small total allocations, the SYCL allocator places multiple buffers on a small subset of channels, causing congestion. At large allocations, the allocator spreads buffers across all channels, reducing contention. The mechanism is shared across multi-channel GPU memory architectures (HBM, GDDR). The methodological implication is \emph{measuring memory bandwidth in isolation systematically underestimates the contention encountered in realistic workloads}, while measuring it co-allocated systematically over-attributes slowdown to the buffer being measured. Both biases are eliminated by the randomized-order methodology.

\subsection{Memory-Subsystem Temporal Warmup}
Cache and TLB warmup behaviour is universal to any GPU with virtual memory and a multi-level cache hierarchy. The phenomenon that the first few memory operations after process start are slower than the steady state is well-known in CPU/GPU benchmarking but rarely quantified at sweep scale. We characterize it on BMG and document the methodological consequences for landscape sweeps.

With isolated single-buffer measurements, read-A time exhibits a 43\% drift across the sequential sweep within each M-block (1024 measurements per M, varying N and K). The drift resets each time M changes.

The Spearman rank correlation of read\_A with sweep run\_order is −0.65 (p \textless{} 0.01, statistically significant negative correlation). However, the intra-process correlation between consecutive measurements is only 0.17 ruling out clock-frequency variation or thermal as the cause.

The GPU\textquotesingle s TLB and L3 cache retain physical-page state across process invocations: after 1000 alloc/free cycles, page-table entries are cached, and address translation becomes faster. The drift is the warm-up curve of the GPU memory pipeline rather than a property of the workload being measured. The same mechanism is present in every GPU memory subsystem we are aware of; the methodological consequence that \textbf{sequential-order landscape sweeps systematically conflate run-order with shape-dependent variables} is what we contribute, along with the randomized-order sweep methodology that resolves it.

\subsection{Resolution: The Randomized-Order Sweep}
To break the spurious correlation between sweep run-order and N (which arises from the nested-loop structure of any sequential sweep), we shuffle all 32,768 (M, N, K) tuples once and run the GPU through 5 warmup invocations before timing.

\includegraphics[width=\linewidth]{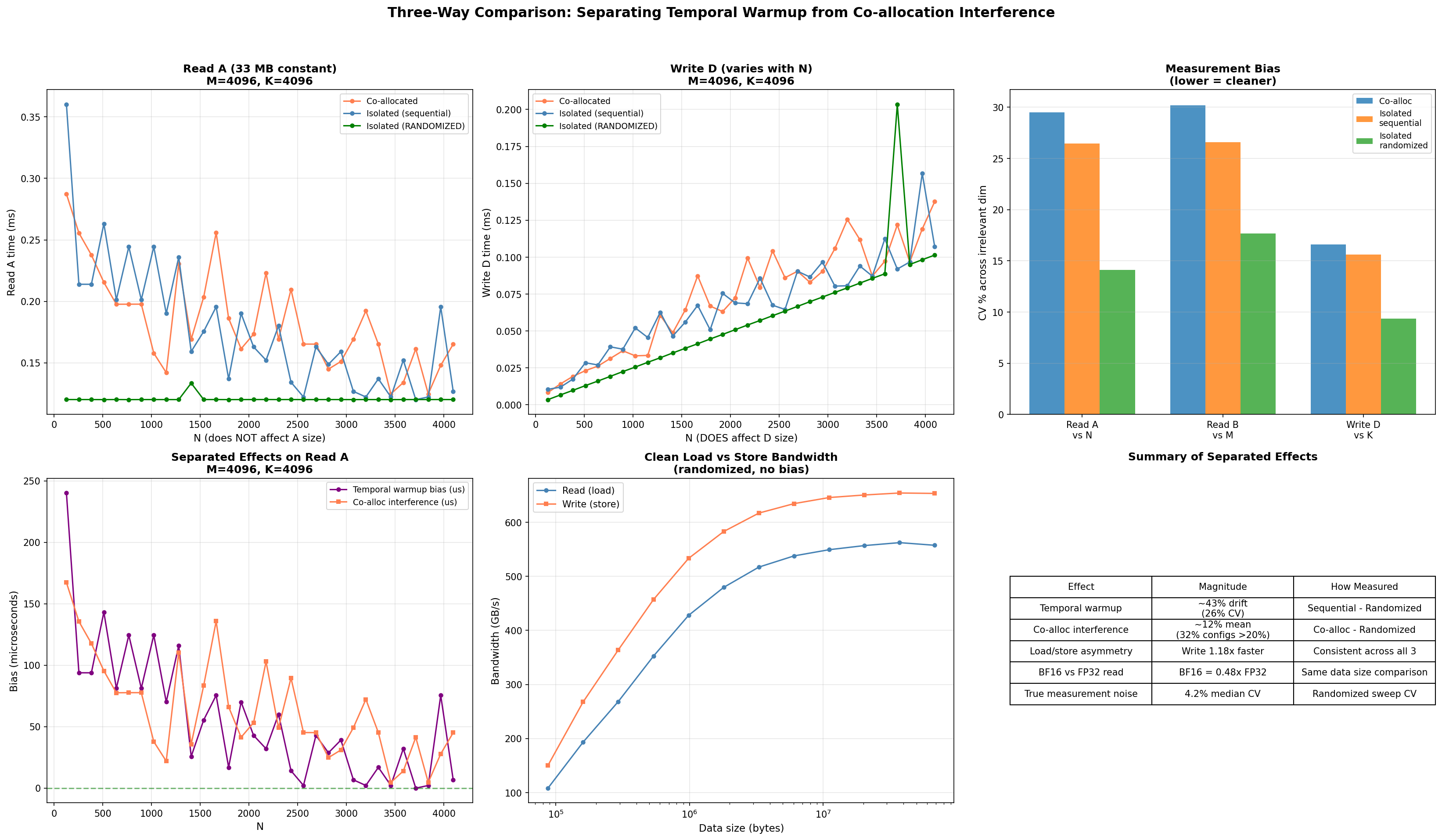}

Figure 9. Three-way comparison of read-A as a function of N. Green (randomized): flat, as expected - N alone has no effect on read-A. Blue (sequential isolated): downward slope - the temporal-warmup artifact. Orange (co-allocated): highest and most variable - temporal warmup combined with co-allocation interference.

The randomized-isolated sweep collapses the spurious corr(read\_A, N) from −0.76 (co-allocated) and −0.66 (sequential) to −0.005, drops the cross-N CV from 30\% / 26\% to 4.2\% median and roughly doubles the apparent achieved bandwidth - read C from 275 to 535 GB/s, write D from 324 to 631 GB/s. (Note: these aggregate values exceed the 456 GB/s GDDR6 spec because the average includes small (M, N, K) configurations whose buffers fit in the 12 MB L3 cache. For configurations with buffers exceeding L3 capacity, achieved bandwidth is bounded by the GDDR6 spec; the cache-amplified averages reflect what GEMM kernels experience for typical-size workloads.)

\begin{longtable}[]{!{\color{black}\vrule}
  >{\raggedright\arraybackslash}p{(\columnwidth - 4\tabcolsep) * \real{0.3347}}!{\color{black}\vrule}
  >{\raggedright\arraybackslash}p{(\columnwidth - 4\tabcolsep) * \real{0.3321}}!{\color{black}\vrule}
  >{\raggedright\arraybackslash}p{(\columnwidth - 4\tabcolsep) * \real{0.3332}}!{\color{black}\vrule}}
\hhline{|-|-|-|}
\cellcolor{tblhdr}\begin{minipage}[b]{\linewidth}\raggedright
\textbf{Effect}
\end{minipage} & \cellcolor{tblhdr}\begin{minipage}[b]{\linewidth}\raggedright
\textbf{Magnitude}
\end{minipage} & \cellcolor{tblhdr}\begin{minipage}[b]{\linewidth}\raggedright
\textbf{How identified}
\end{minipage} \\
\hline
\endhead

\endlastfoot
\textbf{Temporal warmup} & 43\% drift, 26\% CV & Sequential to Randomized \\ \hline
\textbf{Co-allocation interference} & 12\% mean, 32\% configs \textgreater{} 20\% & Co-allocated to isolated \\ \hline
\textbf{Load/Store asymmetry} & Write 1.18X faster & Consistent across 3 sweeps \\ \hline
\textbf{BF16 vs FP32 read benchmark} & BF16 = 0.48 x FP32 BW (per work-item overhead) & Same byte count comparison \\ \hline
\textbf{True measurement noise} & 4.2\% median CV (reduced to 0.05\%) & Within config variance \\ \hline
\end{longtable}

Table 5. Summary of the four memory-subsystem effects identified.

\section{Tile Shape Comparison and Dynamic Selection}

The periodic 256-aligned sawtooth is the dominant feature of the surface texture. Since tile size is a software parameter, the natural question is: how much does varying the tile reduce the sawtooth, and does any single tile dominate?

\subsection{Cross-Tile Setup and Surface Comparison}
We compile six tile variants of the same kernel: 256 × 256 (baseline, 8 × 4 SG layout, 32 SGs / WG), 128 × 128 (4 × 2, 8 SGs), 256 × 128 (8 × 2, 16 SGs), 128 × 256 (4 × 4, 16 SGs), 64 × 64 (4 × 2, 8 SGs), and 64 × 128 (4 × 2, 8 SGs). All use K-tile = 32 and the same DPAS atom; only the workgroup Shape\textless\textgreater{} and subgroup layout differ.

\includegraphics[width=\linewidth]{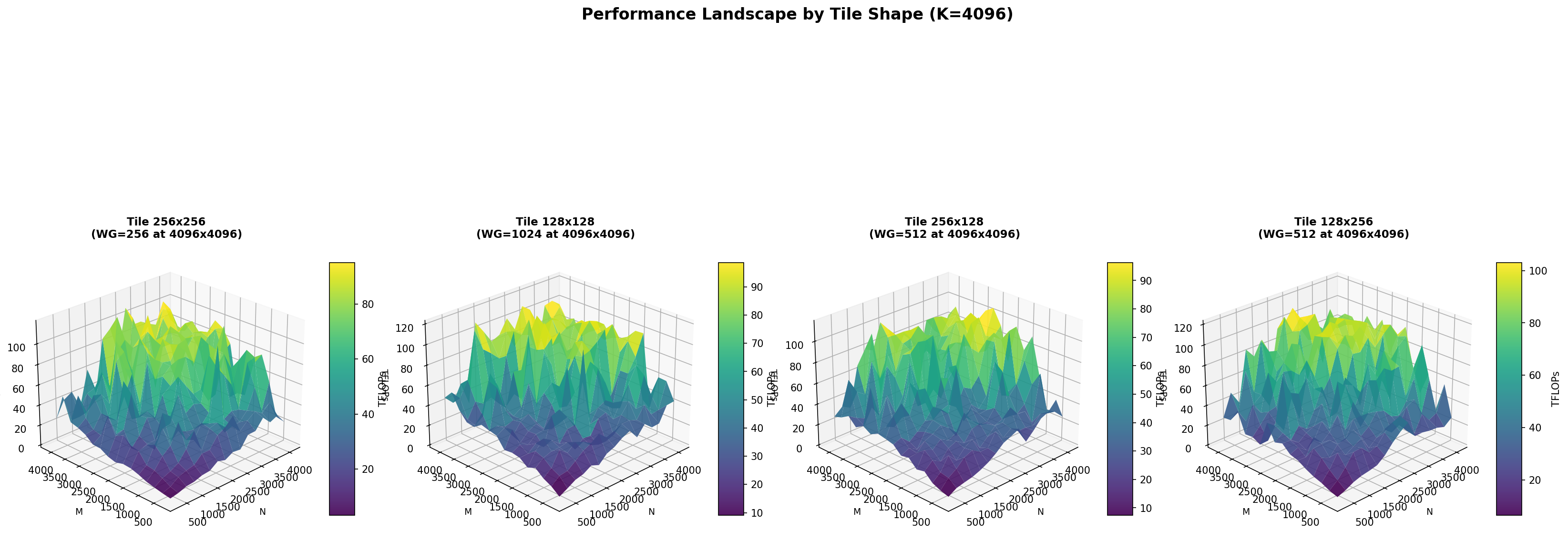}

Figure 10. Performance surfaces for the four primary tiles at K = 4,096. Each tile yields a recognizably different surface shape with the same overall mesa structure but with sawteeth shifted in frequency.

\begin{longtable}[]{!{\color{black}\vrule}
  >{\raggedright\arraybackslash}p{(\columnwidth - 8\tabcolsep) * \real{0.1967}}!{\color{black}\vrule}
  >{\raggedright\arraybackslash}p{(\columnwidth - 8\tabcolsep) * \real{0.1990}}!{\color{black}\vrule}
  >{\raggedright\arraybackslash}p{(\columnwidth - 8\tabcolsep) * \real{0.1990}}!{\color{black}\vrule}
  >{\raggedright\arraybackslash}p{(\columnwidth - 8\tabcolsep) * \real{0.2324}}!{\color{black}\vrule}
  >{\raggedright\arraybackslash}p{(\columnwidth - 8\tabcolsep) * \real{0.1729}}!{\color{black}\vrule}}
\hhline{|-|-|-|-|-|}
\cellcolor{tblhdr}\begin{minipage}[b]{\linewidth}\raggedright
\textbf{Tile}
\end{minipage} & \cellcolor{tblhdr}\begin{minipage}[b]{\linewidth}\raggedright
\textbf{Mean TFLOPs}
\end{minipage} & \cellcolor{tblhdr}\begin{minipage}[b]{\linewidth}\raggedright
\textbf{Max TFLOPs}
\end{minipage} & \cellcolor{tblhdr}\begin{minipage}[b]{\linewidth}\raggedright
\textbf{Peak config}
\end{minipage} & \cellcolor{tblhdr}\begin{minipage}[b]{\linewidth}\raggedright
\textbf{Wins (\% of configs)}
\end{minipage} \\
\hline
\endhead

\endlastfoot
256x256 & 41.2 & 109.0 & 4096x2560x4096 & 16.5\% \\ \hline
128x128 & 44.6 & 112.1 & 3072x2560x3840 & 36.1\% \\ \hline
256x128 & 41.5 & 108.8 & 2560x4096x3840 & 19.3\% \\ \hline
128x256 & 44.5 & 110.2 & 4096x2560x4096 & 28.1\% \\ \hline
\end{longtable}

Table 6. Per-tile aggregate metrics. The 128×128 tile wins on both mean and peak - its smaller per-WG footprint reduces both partial-tile waste and barrier synchronization cost.

\subsection{No Single Tile Is Universally Best}
\includegraphics[width=\linewidth]{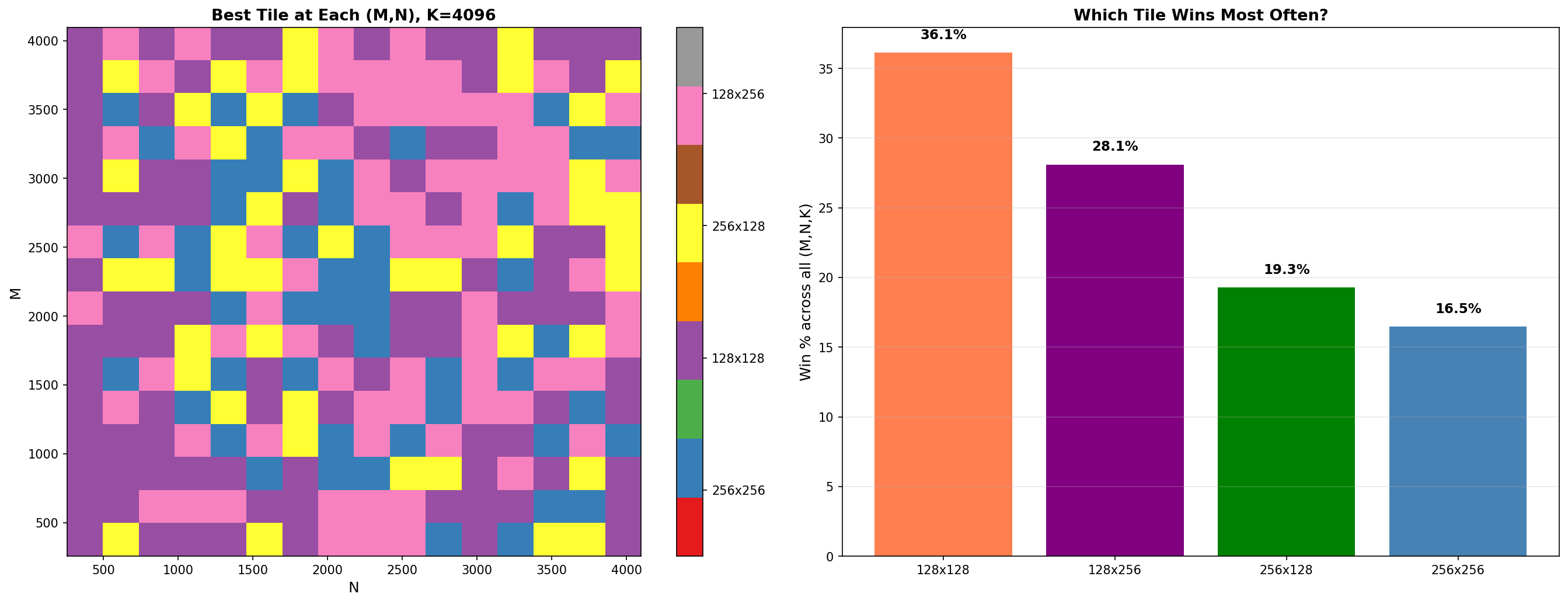}

Figure 11. Winner-tile map at K=4096. The mosaic shows that different (M, N) regions favour different tiles - confirming that runtime tile selection should yield more than any fixed choice.

\subsection{The Sawtooth Tracks the Tile Size}
\includegraphics[width=\linewidth]{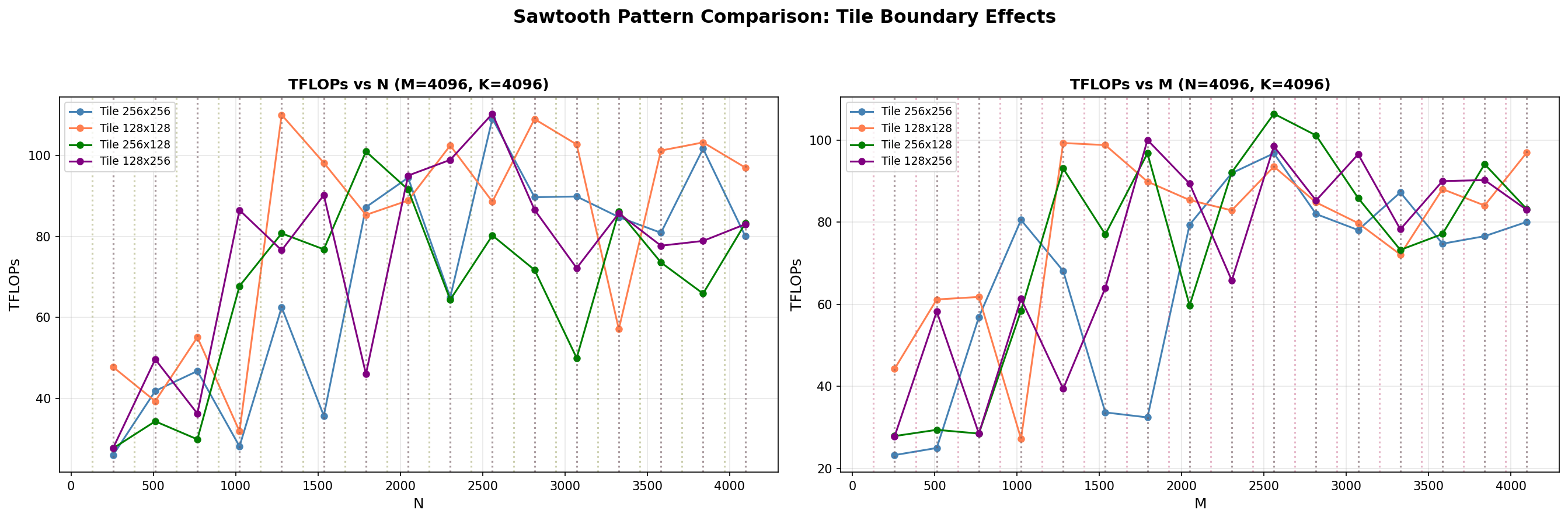}

Figure 12. TFLOPs along varying N (M, K = 4 096) for three tile sizes. The sawtooth period is exactly the tile width: 256 × 256 has period-256, 128 × 128 has period-128, 64 × 64 has period-64. Smaller tiles exhibit smaller amplitude sawteeth because less compute is wasted per partial tile.

The sawtooth period is determined by software, not by hardware cache structure.

\subsection{Dynamic Best-of-Six Tile Selection}
\includegraphics[width=\linewidth]{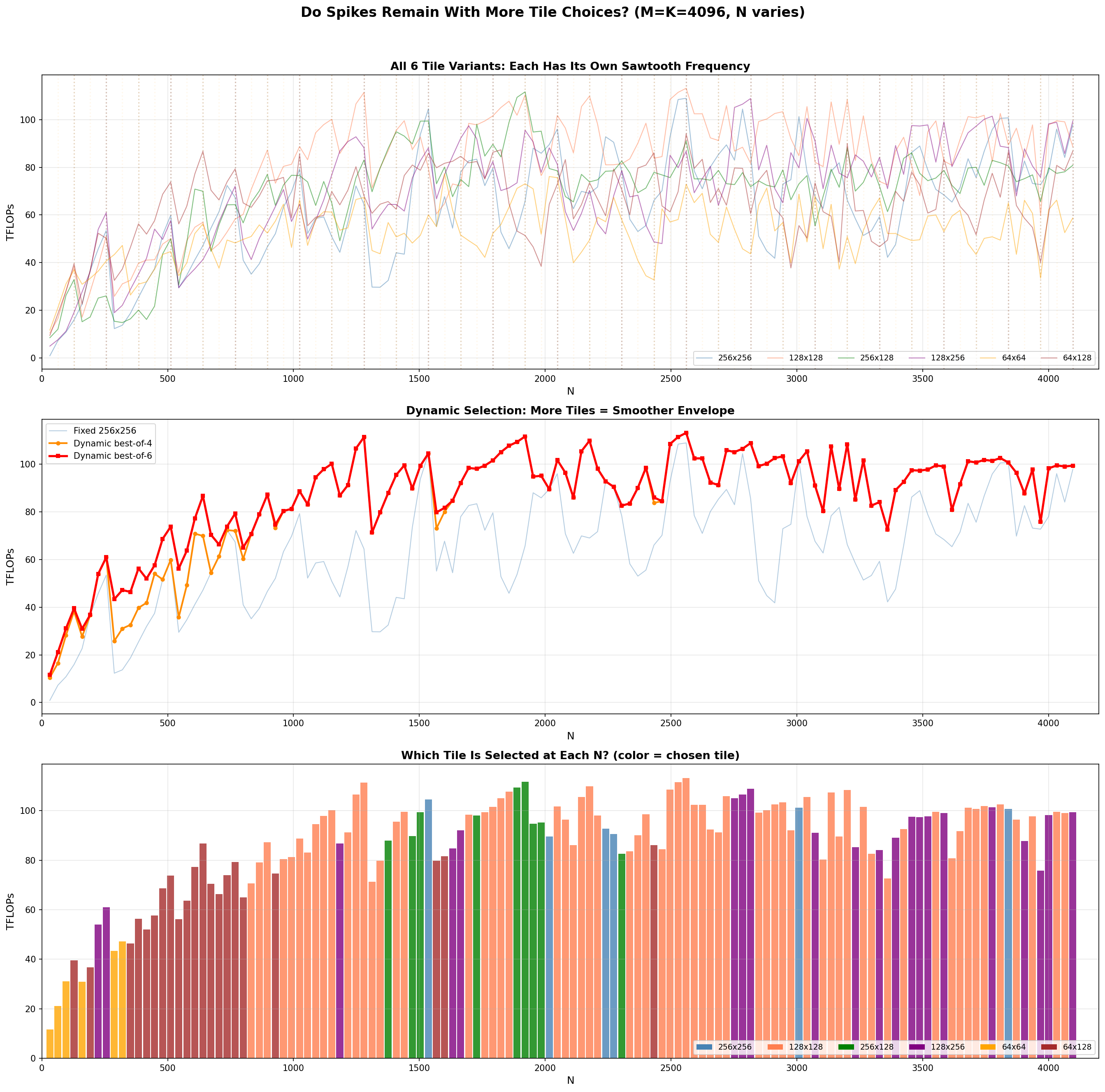}

Figure 13. Fine-grained 1D sweep (M, K = 4 096; N varying step=32) comparing fixed 256 × 256 vs dynamic best-of-4 vs dynamic best-of-6. The best-of-6 envelope is consistently above the fixed tile mean by \textasciitilde35\%.

\begin{longtable}[]{!{\color{black}\vrule}
  >{\raggedright\arraybackslash}p{(\columnwidth - 6\tabcolsep) * \real{0.2545}}!{\color{black}\vrule}
  >{\raggedright\arraybackslash}p{(\columnwidth - 6\tabcolsep) * \real{0.2492}}!{\color{black}\vrule}
  >{\raggedright\arraybackslash}p{(\columnwidth - 6\tabcolsep) * \real{0.2507}}!{\color{black}\vrule}
  >{\raggedright\arraybackslash}p{(\columnwidth - 6\tabcolsep) * \real{0.2457}}!{\color{black}\vrule}}
\hhline{|-|-|-|-|}
\cellcolor{tblhdr}\begin{minipage}[b]{\linewidth}\raggedright
\textbf{Configuration}
\end{minipage} & \cellcolor{tblhdr}\begin{minipage}[b]{\linewidth}\raggedright
\textbf{Mean TFLOP/s}
\end{minipage} & \cellcolor{tblhdr}\begin{minipage}[b]{\linewidth}\raggedright
\textbf{Roughness}
\end{minipage} & \cellcolor{tblhdr}\begin{minipage}[b]{\linewidth}\raggedright
\textbf{CV}
\end{minipage} \\
\hline
\endhead

\endlastfoot
\textbf{Fixed 256x256} & 63.8 & 11.1 & High \\ \hline
\textbf{Dynamic best of 4} & 85.4 & 8.5 & 26.6\% \\ \hline
\textbf{Dynamic best of 6} & 87.1 & 8.0 & 23.1\% \\ \hline
\end{longtable}

Table 7. Aggregate metrics over the fine-N sweep.

Dynamic best-of-6 selection reduces roughness by \textasciitilde28\% over the fixed tile. Across the full 3D landscape, this corresponds to mean TFLOPs 41.2 to 53.9 (+31\%) and roughness 13.5 → 9.3 TFLOPs / step (−31\%) - a substantial smoothing, but the tile-boundary sawteeth are not fully eliminated.

The remaining roughness is the target of the DP pad/split optimization.

We chose six tile variants representing a saturation point in the cost/benefit curve: best-of-4 already captures \textasciitilde96\% of the achievable mean TFLOPs from dynamic tile selection (85.4 vs. 87.1 with best-of-6, Table 7), and additional tile variants have geometrically diminishing returns. We also tested smaller tiles (64×64, 64×128) and observed a crossover: smaller tiles reduce partial-tile waste but increase per-workgroup launch overhead, barrier-synchronization cost and prefetch-pipeline startup overhead. The 64×64 tile, for example, achieves only 46.26 mean TFLOPs at M=K=4096 vs. 87.1 for best-of-6 (Table 13), confirming that the crossover lies near 128×128 for our problem range. The DP padding-and-splitting optimizer (Section 7) reaches the residual roughness floor (\textasciitilde1.31 TFLOPs/step at fine grid, Table 14) more efficiently than further tile proliferation.

\section{Dynamic-Programming based Padding and Splitting}

Tile selection reduces but does not eliminate the tile-boundary sawtooth. We designed two further novel software techniques to address this:

\begin{itemize}
\item
  \textbf{padding-up to a faster nearby shape}, paying for the extra hardware work in exchange for elimination of partial-tile waste; and
\item
  \textbf{splitting a problem into multiple sub-problems} whose total time is less than the original.
\end{itemize}

\begin{quote}
We propose a dynamic-programming algorithm that optimally combines both.
\end{quote}

\subsection{The T0 → T1 → T2 Algorithm}
GEMM problem size is defined as (M, N, K) where matrix A has size (M, K) and matrix B has size (K, N). The output size will be (M, N). We define 3 3-dimensional-data-structures:

\begin{itemize}
\item
  \textbf{T0{[}M{]}{[}N{]}{[}K{]}} = baseline execution time for GEMM (M, N, K) on GPU
\item
  \textbf{T1{[}M{]}{[}N{]}{[}K{]}} = execution time for GEMM of size (M, N, K) on GPU when we can increase the problem size (M, N, K).
\item
  \textbf{T2{[}M{]}{[}N{]}{[}K{]}} = execution time for GEMM of problem size (M, N, K) on GPU when we can increase or split the problem size or both.
\end{itemize}

Computing these 3 data-structures is done in an offline auto-tuning phase which is done once for a given hardware-software stack combination. The final data structure T2{[}M{]}{[}N{]}{[}K{]} is used to split or increase input matrix size during runtime in O(1) time complexity. There are 3 core steps:

\begin{itemize}
\item
  Step 1: Benchmarking GEMM / Compute T0
\item
  Step 2: Compute T1: Increase matrix size
\item
  Step 3: Compute T2: Split + increase matrix size
\end{itemize}

\textbf{Step 1: Benchmarking GEMM / Compute T0}

Capture all values of T0{[}M{]}{[}N{]}{[}K{]}: Benchmark on the actual GPU hardware with the targeted software stack.

\textbf{Step 2: Compute T1: Increase matrix size}

T1 is computed in (Bottom Right to Top Left) approach. T1 is initialized as T0.

\[T_{1}\lbrack M\rbrack\lbrack N\rbrack\lbrack K\rbrack = \min_{(i,j,k) \in \text{\{}0,1\text{\}}^{3}}T_{1}\lbrack M + i\rbrack\lbrack N + j\rbrack\lbrack K + k\rbrack\]

Where \((i,j,k)\) can be 0 or 1. This is computed in \(O(MNK)\) time complexity.

\textbf{Step 3: Compute T2: Split + increase matrix size}

T2 is computed in reverse order as of T1 that is (Top left to Bottom right) approach. T2 is initialized as T1. To compute T2{[}M{]}{[}N{]}{[}K{]}, we used values of T2{[}M1{]}{[}N1{]}{[}K1{]} which are already computed where:

\begin{itemize}
\item
  M1 \textless{} M
\item
  N1 \textless{} N
\item
  K1 \textless{} K
\end{itemize}

Formula to compute T2:

\begin{verbatim}
T2[M][N][K] = MINIMUM( T2[M][N][K],
                       MINIMUM( T2[M-i][N][K] + T2[i][N][K] ),
                       MINIMUM( T2[M][N-j][K] + T2[M][j][K] ),
                       MINIMUM( T2[M][N][K-k] + T2[M][N][k] ) )
\end{verbatim}

Time complexity: \(O\left( (MNK)*(M + N + K) \right)\)

For simplicity, if N=M=K, then time complexity to compute T2 is O(N\textsuperscript{4}).

Explanation of the input matrix split:

\textbf{Split on M dimension} only:

\begin{itemize}
\item
  Matrix A is split. Output is concatenated.
\end{itemize}

\includegraphics[width=\linewidth]{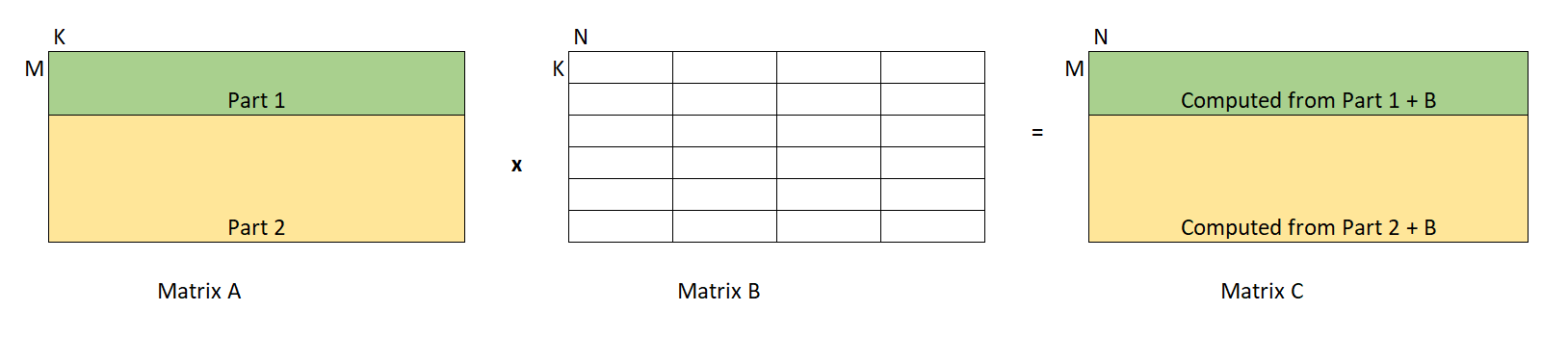}

Figure 14. Split matrix A row-wise.

\textbf{Split on N dimension} only:

\begin{itemize}
\item
  Matrix B is split. Output is concatenated.
\end{itemize}

\includegraphics[width=\linewidth]{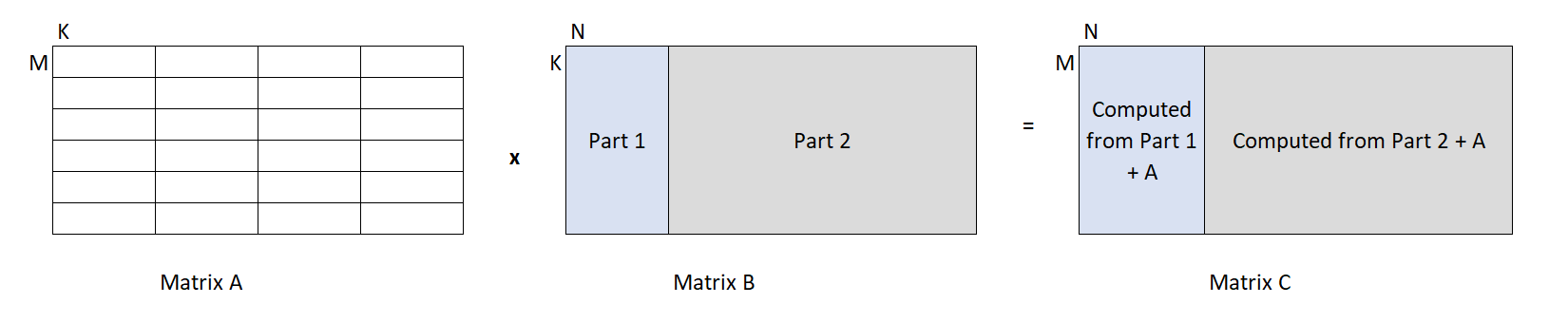}

Figure 15. Split matrix B column-wise.

\textbf{Split on K dimension} only:

\begin{itemize}
\item
  Both matrix A and B are split.
\item
  Results in 2 partial outputs which are added elementwise to get the final output.
\item
  GEMM for the first part is computed first (Part 11 + 21). This output can be passed as C matrix with beta=1 for the second GEMM. This has negligible overhead as the elementwise addition is fused with GEMM kernel in practice.
\item
  Both GEMMs are computed sequentially on the same GPU. Overall execution time is improved.
\end{itemize}

\includegraphics[width=\linewidth]{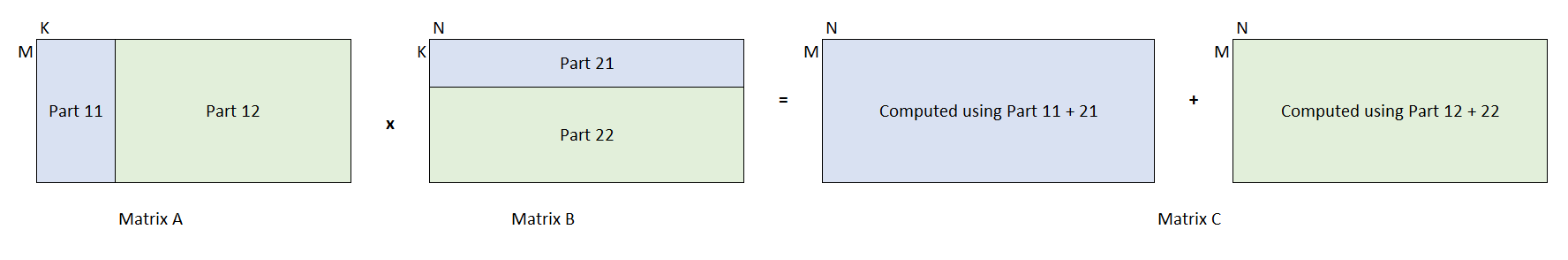}

Figure 16. Split matrix A column-wise and matrix B row-wise.

\textbf{Complexity}: For an \(NxNxN\) grid, both T₁ and T₂ are computable in \textbf{O(N⁴)}. The final T₂ table provides O(1) runtime lookup of the optimal padding-or-splitting decision for every (M, N, K).

\subsection{Results on Fixed Tile (256 × 256; 32,768 Configurations)}
\includegraphics[width=\linewidth]{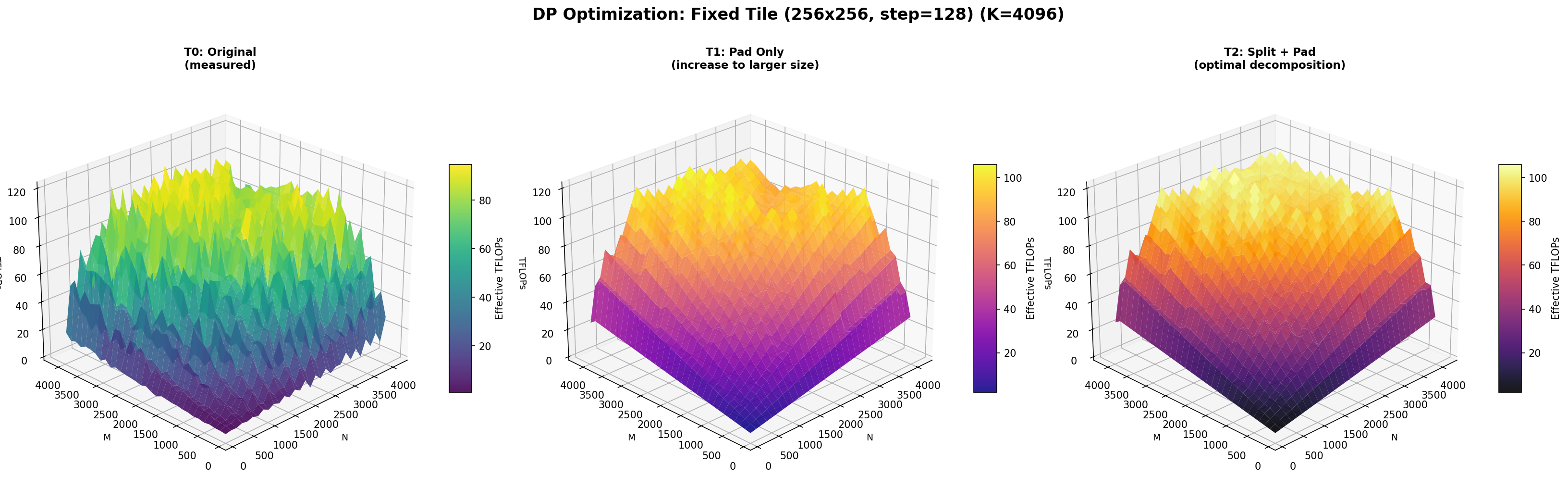}

Figure 17. Three surfaces at K=4096. Left (T0): jagged original landscape with deep valleys. Middle (T1): padding alone fills most valleys; the surface is much smoother. Right (T2): split + pad produces the smoothest surface with the highest floor.

\includegraphics[width=\linewidth]{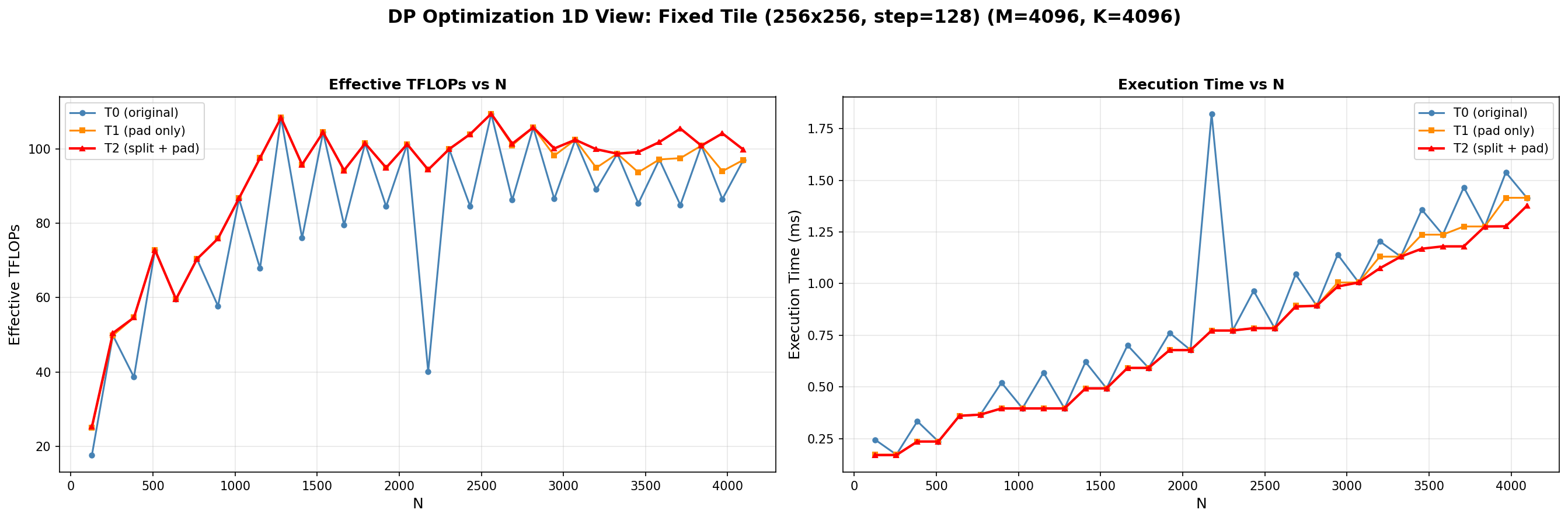}

Figure 18. 1D slice (M, K = 4096; varying N). Wild oscillations in T0 (blue) collapse to a steady 70 to 100+ TFLOPs in T1 (orange). Splitting (T2, red) provides additional smoothing especially at large N.

\begin{longtable}[]{!{\color{black}\vrule}
  >{\raggedright\arraybackslash}p{(\columnwidth - 4\tabcolsep) * \real{0.3361}}!{\color{black}\vrule}
  >{\raggedright\arraybackslash}p{(\columnwidth - 4\tabcolsep) * \real{0.3319}}!{\color{black}\vrule}
  >{\raggedright\arraybackslash}p{(\columnwidth - 4\tabcolsep) * \real{0.3320}}!{\color{black}\vrule}}
\hhline{|-|-|-|}
\cellcolor{tblhdr}\begin{minipage}[b]{\linewidth}\raggedright
\textbf{Metric}
\end{minipage} & \cellcolor{tblhdr}\begin{minipage}[b]{\linewidth}\raggedright
\textbf{T0 to T1 (pad)}
\end{minipage} & \cellcolor{tblhdr}\begin{minipage}[b]{\linewidth}\raggedright
\textbf{T0 to T2 (split + pad)}
\end{minipage} \\
\hline
\endhead

\endlastfoot
\textbf{Mean time reduction} & 22.7\% & \textbf{24.4\%} \\ \hline
\textbf{Max time reduction} & 87.6\% & 87.6\% \\ \hline
\textbf{Configs improved \textgreater{} 10\%} & 74\% & 74\% \\ \hline
\textbf{Configs improved \textgreater{} 20\%} & 53\% & 53\% \\ \hline
\textbf{Surface roughness reduction} & 66.7\% & \textbf{68.2\%} \\ \hline
\end{longtable}

Table 8. DP impact on the fixed-tile landscape.

\subsection{What Does the DP Choose?}
\begin{longtable}[]{!{\color{black}\vrule}
  >{\raggedright\arraybackslash}p{(\columnwidth - 4\tabcolsep) * \real{0.3321}}!{\color{black}\vrule}
  >{\raggedright\arraybackslash}p{(\columnwidth - 4\tabcolsep) * \real{0.2576}}!{\color{black}\vrule}
  >{\raggedright\arraybackslash}p{(\columnwidth - 4\tabcolsep) * \real{0.4102}}!{\color{black}\vrule}}
\hhline{|-|-|-|}
\cellcolor{tblhdr}\begin{minipage}[b]{\linewidth}\raggedright
\textbf{Action}
\end{minipage} & \cellcolor{tblhdr}\begin{minipage}[b]{\linewidth}\raggedright
\textbf{Fraction at K=4096}
\end{minipage} & \cellcolor{tblhdr}\begin{minipage}[b]{\linewidth}\raggedright
\textbf{Where it appears}
\end{minipage} \\
\hline
\endhead

\endlastfoot
\textbf{Pad or as-is (no split)} & 75\% & Pervasive \\ \hline
\textbf{Split on K} & 13\% & Large M, N (compute-bound region) \\ \hline
\textbf{Split on N} & 8\% & At specific ``bad N'' values \\ \hline
\textbf{Split on M} & 5\% & Edge configurations \\ \hline
\end{longtable}

Table 9. Action distribution chosen by the DP at K = 4096.

K-splitting is most common at large M, N where compute-bound problems benefit from exposing more parallelism via split-K accumulation. N-splitting targets specific bad-N values where cache or layout effects make any single kernel slow. M-splitting is rare.

\subsection{Results on the Dynamic-Tile Landscape}
Applied on top of the dynamic-tile baseline, the DP still finds 10.4\% mean time reduction (T1) and 11.6\% (T2), with 27.8\% additional surface-roughness reduction. The fact that DP improvement persists on top of dynamic tile selection confirms that padding and splitting address a different source of variation than tile choice does.

\subsection{The Combined Optimization Stack}
\includegraphics[width=\linewidth]{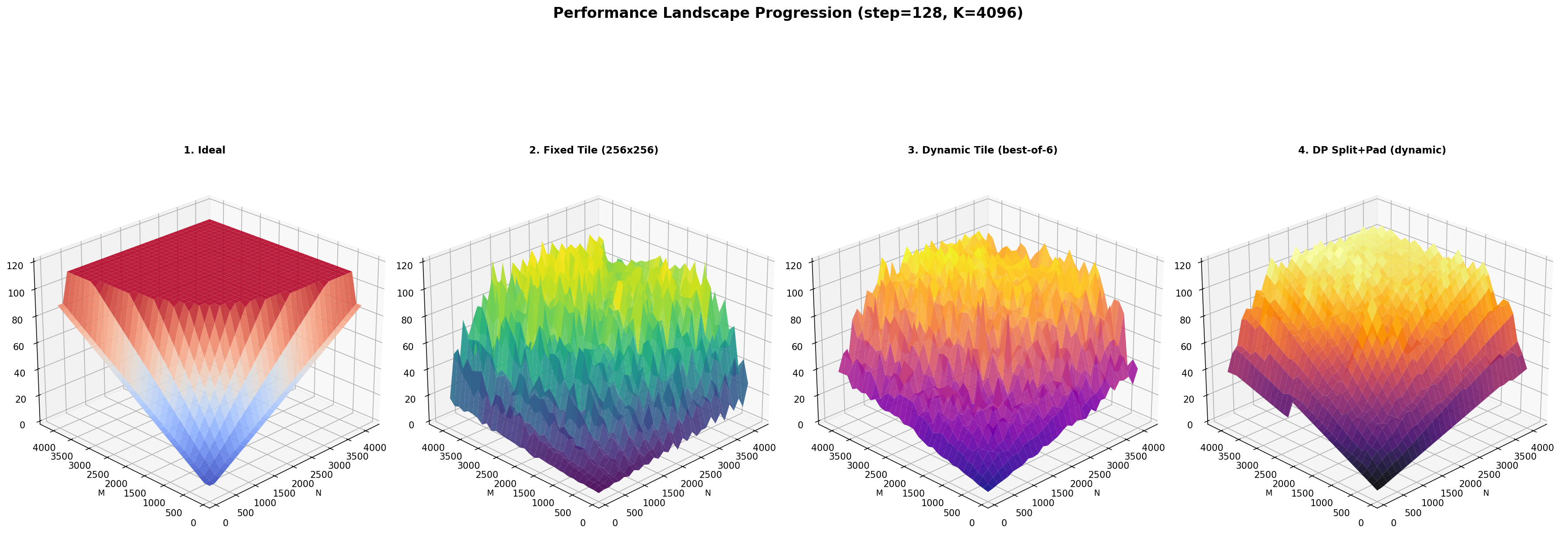}

\includegraphics[width=\linewidth]{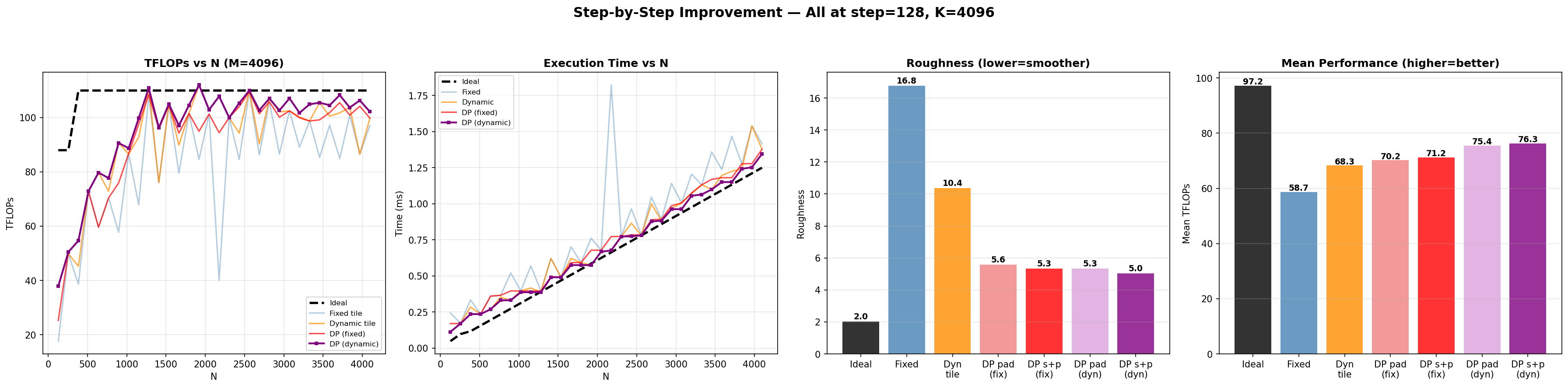}

Figure 19. Five-stage progression of the landscape at uniform step=128 resolution at K = 4 096: ideal to fixed tile to dynamic tile to DP-pad on dynamic to DP split+pad on dynamic.

\begin{longtable}[]{!{\color{black}\vrule}
  >{\raggedright\arraybackslash}p{(\columnwidth - 6\tabcolsep) * \real{0.3066}}!{\color{black}\vrule}
  >{\raggedright\arraybackslash}p{(\columnwidth - 6\tabcolsep) * \real{0.1934}}!{\color{black}\vrule}
  >{\raggedright\arraybackslash}p{(\columnwidth - 6\tabcolsep) * \real{0.2395}}!{\color{black}\vrule}
  >{\raggedright\arraybackslash}p{(\columnwidth - 6\tabcolsep) * \real{0.2605}}!{\color{black}\vrule}}
\hhline{|-|-|-|-|}
\cellcolor{tblhdr}\begin{minipage}[b]{\linewidth}\raggedright
\textbf{Stage}
\end{minipage} & \cellcolor{tblhdr}\begin{minipage}[b]{\linewidth}\raggedright
\textbf{Mean TFLOP/s}
\end{minipage} & \cellcolor{tblhdr}\begin{minipage}[b]{\linewidth}\raggedright
\textbf{Roughness (TFLOPs/step)}
\end{minipage} & \cellcolor{tblhdr}\begin{minipage}[b]{\linewidth}\raggedright
\textbf{Vs Ideal roughness}
\end{minipage} \\
\hline
\endhead

\endlastfoot
\textbf{Ideal (compute-only)} & \textbf{97.2} & \textbf{2.0} & \textbf{1.0x} \\ \hline
\textbf{Fixed tile (256x256)} & \textbf{58.7} & \textbf{16.8} & \textbf{8.2x} \\ \hline
\textbf{Dynamic tile (best of 6)} & 68.3 & 10.4 & 5.1x \\ \hline
\textbf{DP pad (fixed tile)} & 70.2 & 5.6 & 2.7x \\ \hline
\textbf{DP split+pad (fixed tile)} & 71.2 & 5.3 & 2.6x \\ \hline
\textbf{DP pad (dynamic tile)} & 75.4 & 5.3 & 2.6x \\ \hline
\textbf{DP split+pad (dynamic tile)} & \textbf{76.3} & \textbf{5.0} & \textbf{2.5x} \\ \hline
\end{longtable}

Table 10. End-to-end optimization stack.

Mean TFLOPs improves monotonically; roughness drops monotonically. The best stack achieves +30\% mean and -70\% roughness vs the fixed-tile baseline. The remaining 2.5 gap to the ideal roughness floor is the target of the residual analysis.

\section{Mechanism Refinement: Four Targeted Experiments}

After the optimization stack reduces landscape roughness from 16.8 to 5.0 TFLOPs / step (against an ideal of 2.0), one question remains: \textbf{what is the residual 2.5x gap?}

With four targeted experiments, we identified that the periodic structure of the residual is software-mechanical (partial-tile waste at progressively finer scales) while the non-periodic component is genuinely hardware-bound.

\subsection{Independent Measurement on Hand-Picked Configurations}
To verify that the suspected tile-misaligned configurations are reproducibly slower (rather than measurement noise), we extract per-kernel GPU execution times for 20 hand-picked configurations grouped into four categories:

\begin{itemize}
\item
  tile-aligned (N ∈ \{1024, 2048, 4096\} at M = 4096 plus 3840 × 2048 and 2048 × 2048)
\item
  tile-misaligned at M = 4096 (N ∈ \{3168, 3200, 3232, 3264, 3968\})
\item
  fixed-N varying-M control group (N = 3168 with M ∈ \{1024, 2048, 3072, 4096\}, plus N = 3232 with M = 2048)
\item
  saturation-curve transitions (M × N ∈ \{256\textsuperscript{2}, 512\textsuperscript{2}, 1024\textsuperscript{2}, 2560 × 1536, 3840 × 2048\})
\end{itemize}

\begin{quote}
Per-kernel timing was extracted via Intel VTune\textquotesingle s gpu-hotspots profiler {[}9{]} (kernel attribution mode) which provides \textasciitilde0.1\% timing precision through direct hardware kernel-launch instrumentation.
\end{quote}

\begin{longtable}[]{!{\color{black}\vrule}
  >{\raggedright\arraybackslash}p{(\columnwidth - 4\tabcolsep) * \real{0.3351}}!{\color{black}\vrule}
  >{\raggedright\arraybackslash}p{(\columnwidth - 4\tabcolsep) * \real{0.3338}}!{\color{black}\vrule}
  >{\raggedright\arraybackslash}p{(\columnwidth - 4\tabcolsep) * \real{0.3312}}!{\color{black}\vrule}}
\hhline{|-|-|-|}
\cellcolor{tblhdr}\begin{minipage}[b]{\linewidth}\raggedright
\textbf{Group}
\end{minipage} & \cellcolor{tblhdr}\begin{minipage}[b]{\linewidth}\raggedright
\textbf{Mean TFLOPs}
\end{minipage} & \cellcolor{tblhdr}\begin{minipage}[b]{\linewidth}\raggedright
\textbf{STD}
\end{minipage} \\
\hline
\endhead

\endlastfoot
Tile-aligned & 95.19 & 6.92 \\ \hline
Tile-misaligned & 86.40 & 4.38 \\ \hline
Fixed-N varying-M control & 75.71 & 4.69 \\ \hline
Saturation transitions & 66.48 & 43.99 \\ \hline
\end{longtable}

Table 11. Independently measured kernel-time TFLOPs by group.

Head-to-head at M = 4096: tile-aligned 93.65 vs tile-misaligned 86.40 - a reproducible 7.7\% slowdown. The fixed-N varying-M control group shows that the slowdown is intrinsic to N and not an M × N interaction effect, and the saturation group spans the expected mesa shape from launch-dominated through plateau.

\subsection{Per-Configuration Timing Variance}
To check whether ruggedness is ``noise'' or deterministic signal, we run each of five representative configurations 50 times as independent process invocations.

\begin{longtable}[]{!{\color{black}\vrule}
  >{\raggedright\arraybackslash}p{(\columnwidth - 8\tabcolsep) * \real{0.5231}}!{\color{black}\vrule}
  >{\raggedright\arraybackslash}p{(\columnwidth - 8\tabcolsep) * \real{0.1832}}!{\color{black}\vrule}
  >{\raggedright\arraybackslash}p{(\columnwidth - 8\tabcolsep) * \real{0.0999}}!{\color{black}\vrule}
  >{\raggedright\arraybackslash}p{(\columnwidth - 8\tabcolsep) * \real{0.1000}}!{\color{black}\vrule}
  >{\raggedright\arraybackslash}p{(\columnwidth - 8\tabcolsep) * \real{0.0939}}!{\color{black}\vrule}}
\hhline{|-|-|-|-|-|}
\cellcolor{tblhdr}\begin{minipage}[b]{\linewidth}\raggedright
\textbf{Configuration}
\end{minipage} & \cellcolor{tblhdr}\begin{minipage}[b]{\linewidth}\raggedright
\textbf{Mean TFLOPs}
\end{minipage} & \cellcolor{tblhdr}\begin{minipage}[b]{\linewidth}\raggedright
\textbf{STD}
\end{minipage} & \cellcolor{tblhdr}\begin{minipage}[b]{\linewidth}\raggedright
\textbf{CV}
\end{minipage} & \cellcolor{tblhdr}\begin{minipage}[b]{\linewidth}\raggedright
\textbf{Range}
\end{minipage} \\
\hline
\endhead

\endlastfoot
\textbf{Tile-aligned, rectangular (4096 × 2048 × 4096)} & 99.26 & 0.069 & 0.07\% & 0.29\% \\ \hline
\textbf{Tile-aligned, square (4096 × 4096 × 4096)} & 96.22 & 0.055 & 0.06\% & 0.26\% \\ \hline
\textbf{Tile-misaligned, N = 3168 (4096 × 3168 × 4096)} & 81.77 & 0.033 & 0.04\% & 0.21\% \\ \hline
\textbf{Tile-misaligned, N = 3096 (4096 × 3096 × 4096)} & 70.51 & 0.069 & 0.10\% & 0.49\% \\ \hline
\textbf{Saturation (2560 × 1536 × 4096)} & 109.55 & 0.063 & 0.06\% & 0.31\% \\ \hline
\end{longtable}

Table 12. Within-configuration variance across 50 independent runs.

The within-configuration CV is 0.04 to 0.10\% - the GPU is essentially deterministic. The ``4.2\% within-plateau CV'' from the randomized sweep is not measurement noise but real configuration-dependent variation. Every visible feature of the rugged landscape is deterministic signal that depends only on (M, N, K).

\subsection{Cross-Tile Fine-N Sweep: The Definitive Mechanism Test}
We sweep N from 3,000 to 4,096 in steps of 32 at fixed M=K=4096, for three tile sizes (64x64, 128x128, 256x256). If the residual ruggedness arose from L3 cache set conflicts, valleys would appear at the same N values regardless of tile choice. They do not.

\includegraphics[width=\linewidth]{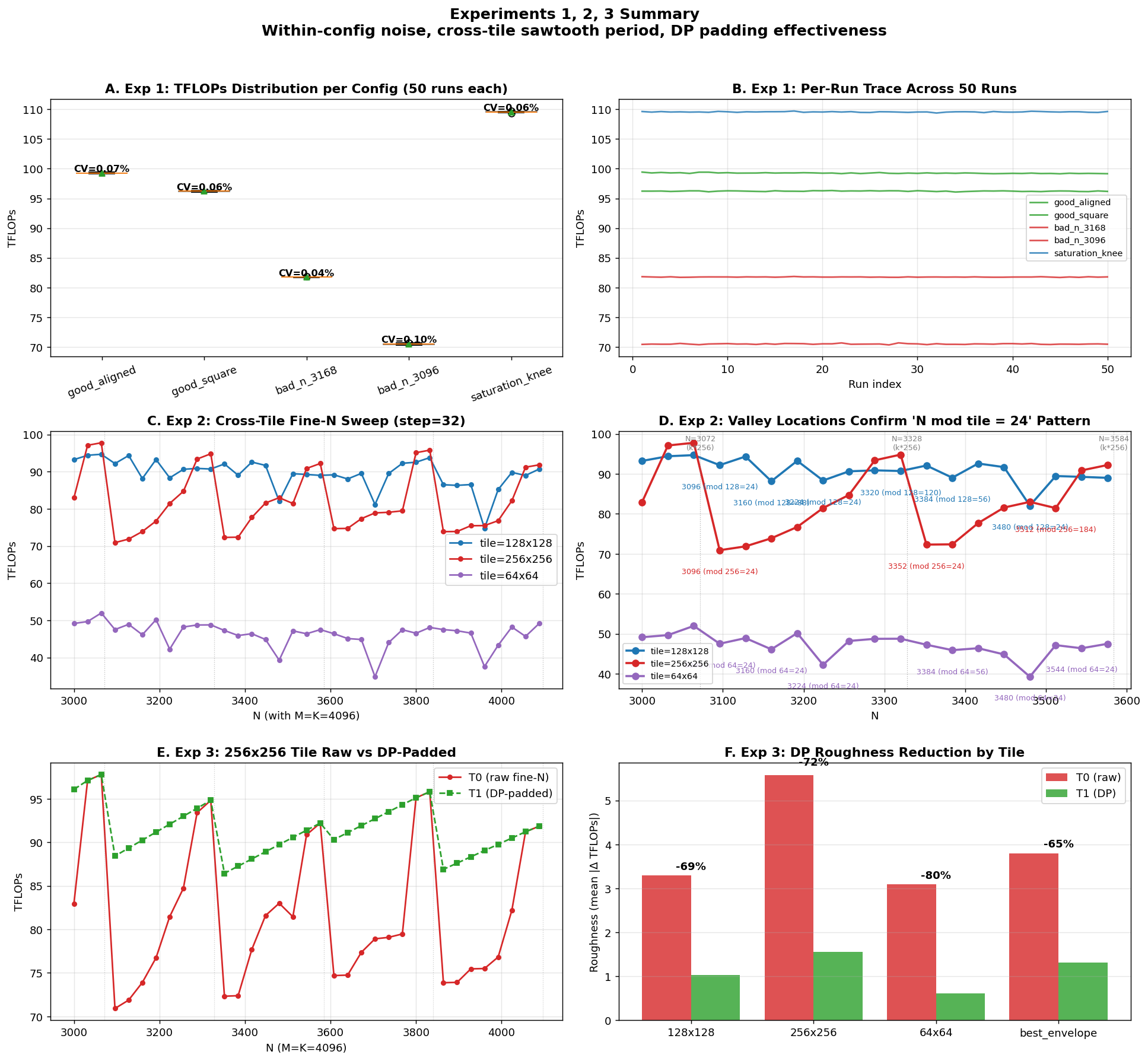}

Figure 20. Combined view of all four follow-up experiments.

Panel A: per-config CV bars from Exp 1 (all CVs ≤ 0.10\%). Panel B: the same shows trace overlay across 50 runs for each configuration - flat lines, confirming determinism. Panel C: cross-tile fine-N sweep. Panel D: zoom on N=3000 to 3600 with valleys annotated, showing valleys consistently at N mod tile\_size = 24 for all three tiles. Panels E, F: DP padding effect.

Valleys consistently appear at the smallest non-zero N mod tile\_size value reachable in our step-32 sampling (which is 24 for the starting offset N = 3000). The mechanism predicts the deepest valley occurs at N mod tile\_size = 0⁺ (smallest non-zero remainder); finer sampling would reveal proportionally deeper valleys at smaller remainders. The qualitative phenomenon - valley alignment with the smallest non-zero remainder - is robust across all three tile sizes.

\begin{longtable}[]{!{\color{black}\vrule}
  >{\raggedright\arraybackslash}p{(\columnwidth - 6\tabcolsep) * \real{0.1400}}!{\color{black}\vrule}
  >{\raggedright\arraybackslash}p{(\columnwidth - 6\tabcolsep) * \real{0.3165}}!{\color{black}\vrule}
  >{\raggedright\arraybackslash}p{(\columnwidth - 6\tabcolsep) * \real{0.2498}}!{\color{black}\vrule}
  >{\raggedright\arraybackslash}p{(\columnwidth - 6\tabcolsep) * \real{0.2937}}!{\color{black}\vrule}}
\hhline{|-|-|-|-|}
\cellcolor{tblhdr}\begin{minipage}[b]{\linewidth}\raggedright
\textbf{Tile}
\end{minipage} & \cellcolor{tblhdr}\begin{minipage}[b]{\linewidth}\raggedright
\textbf{Median sawtooth period}
\end{minipage} & \cellcolor{tblhdr}\begin{minipage}[b]{\linewidth}\raggedright
\textbf{Valleys}
\end{minipage} & \cellcolor{tblhdr}\begin{minipage}[b]{\linewidth}\raggedright
\textbf{Mean TFLOPs}
\end{minipage} \\
\hline
\endhead

\endlastfoot
\textbf{256x256} & 256 & N mod 256 = 24 & 82.09 \\ \hline
\textbf{128x128} & 128 & N mod 128 = 24 & 89.46 \\ \hline
\textbf{64x64} & 64 & N mod 64 = 24 & 46.26 \\ \hline
\end{longtable}

Table 13. Sawtooth period scales exactly with tile size at all three scales.

If the residual ruggedness arose from L3 cache set conflicts, valleys would appear at the same N values regardless of tile choice - cache-conflict effects depend on the memory-access stride N x sizeof(element), which is fixed by N and independent of the workgroup tile. Partial-tile waste, in contrast, depends on N mod tile\_N and therefore shifts when the tile changes. We observe valleys at N mod tile\_N = 24 for all three tile sizes (Table 13), confirming partial-tile waste rather than cache conflict as the periodic-structure mechanism.

\subsection{DP Padding on the Fine-N Data}
To test how much further DP padding can reach at the fine grid (where the step-128 DP could not act), we apply T0 to T1 separately to each tile\textquotesingle s fine-N sweep. The summary panel for this experiment is panels E and F of Figure 20.

\begin{longtable}[]{!{\color{black}\vrule}
  >{\raggedright\arraybackslash}p{(\columnwidth - 12\tabcolsep) * \real{0.1068}}!{\color{black}\vrule}
  >{\raggedright\arraybackslash}p{(\columnwidth - 12\tabcolsep) * \real{0.1665}}!{\color{black}\vrule}
  >{\raggedright\arraybackslash}p{(\columnwidth - 12\tabcolsep) * \real{0.1618}}!{\color{black}\vrule}
  >{\raggedright\arraybackslash}p{(\columnwidth - 12\tabcolsep) * \real{0.1214}}!{\color{black}\vrule}
  >{\raggedright\arraybackslash}p{(\columnwidth - 12\tabcolsep) * \real{0.1165}}!{\color{black}\vrule}
  >{\raggedright\arraybackslash}p{(\columnwidth - 12\tabcolsep) * \real{0.1499}}!{\color{black}\vrule}
  >{\raggedright\arraybackslash}p{(\columnwidth - 12\tabcolsep) * \real{0.1772}}!{\color{black}\vrule}}
\hhline{|-|-|-|-|-|-|-|}
\cellcolor{tblhdr}\begin{minipage}[b]{\linewidth}\raggedright
\textbf{Tile}
\end{minipage} & \cellcolor{tblhdr}\begin{minipage}[b]{\linewidth}\raggedright
\textbf{T0 roughness}
\end{minipage} & \cellcolor{tblhdr}\begin{minipage}[b]{\linewidth}\raggedright
\textbf{T1 roughness}
\end{minipage} & \cellcolor{tblhdr}\begin{minipage}[b]{\linewidth}\raggedright
\textbf{Reduction}
\end{minipage} & \cellcolor{tblhdr}\begin{minipage}[b]{\linewidth}\raggedright
\textbf{T0 mean}
\end{minipage} & \cellcolor{tblhdr}\begin{minipage}[b]{\linewidth}\raggedright
\textbf{T1 mean}
\end{minipage} & \cellcolor{tblhdr}\begin{minipage}[b]{\linewidth}\raggedright
\textbf{Min: T0 to T1}
\end{minipage} \\
\hline
\endhead

\endlastfoot
\textbf{256x256} & 5.58 & 1.56 & 72\% & 82.09 & 91.39 & 70.92 to 86.40 \\ \hline
\textbf{128x128} & 3.30 & 1.03 & 69\% & 89.46 & 90.90 & 74.76 to 86.50 \\ \hline
\textbf{64x64} & 3.09 & 0.61 & 80\% & 46.26 & 47.99 & 34.84 to 45.92 \\ \hline
\textbf{Best of envelope} & 3.79 & 1.31 & 65\% & 90.23 & 92.29 & 75.50 to 86.90 \\ \hline
\end{longtable}

Table 14. DP padding eliminates 65 to 80\% of the residual sawtooth across all tiles.

The remaining roughness of \textasciitilde1.31 TFLOPs / step is approximately 22x the within-config measurement noise (0.06 TFLOPs from Table 12). It is real signal, but its magnitude is only \textasciitilde1.5\% of mean TFLOPs.

\begin{longtable}[]{!{\color{black}\vrule}
  >{\raggedright\arraybackslash}p{(\columnwidth - 8\tabcolsep) * \real{0.1532}}!{\color{black}\vrule}
  >{\raggedright\arraybackslash}p{(\columnwidth - 8\tabcolsep) * \real{0.3075}}!{\color{black}\vrule}
  >{\raggedright\arraybackslash}p{(\columnwidth - 8\tabcolsep) * \real{0.1691}}!{\color{black}\vrule}
  >{\raggedright\arraybackslash}p{(\columnwidth - 8\tabcolsep) * \real{0.2012}}!{\color{black}\vrule}
  >{\raggedright\arraybackslash}p{(\columnwidth - 8\tabcolsep) * \real{0.1691}}!{\color{black}\vrule}}
\hhline{|-|-|-|-|-|}
\cellcolor{tblhdr}\begin{minipage}[b]{\linewidth}\raggedright
\textbf{Source}
\end{minipage} & \cellcolor{tblhdr}\begin{minipage}[b]{\linewidth}\raggedright
\textbf{Active criterion}
\end{minipage} & \cellcolor{tblhdr}\begin{minipage}[b]{\linewidth}\raggedright
\textbf{Active fraction}
\end{minipage} & \cellcolor{tblhdr}\begin{minipage}[b]{\linewidth}\raggedright
\textbf{Per-event swing}
\end{minipage} & \cellcolor{tblhdr}\begin{minipage}[b]{\linewidth}\raggedright
\textbf{Contribution (TFLOPs/step)}
\end{minipage} \\
\hline
\endhead

\endlastfoot
\textbf{Wave-fill effects on 20 Xe-cores} & Wave count V = ⌈W/20⌉ changes between adjacent sweep points (captures both monotone ramp from partial-wave underutilization at small problems and tail-effect oscillation as W crosses wave boundaries) & \textasciitilde100\% for ramp-up + \textasciitilde20\% for wave underutilization & \textasciitilde6 TFLOPs (2 +4) & \textasciitilde2.8 \\ \hline
\textbf{Per-kernel overhead variation} & Every kernel launch incurs \textasciitilde32\% mean overhead (barrier latency + register spills + GDDR6 latency); shape-dependent variation across (M, N, K) & 100\% & \textasciitilde1.5 TFLOPs & \textasciitilde1.5 \\ \hline
\textbf{DPAS atom M:N asymmetry residual} & Residual N-cliff vs M-cliff asymmetry after padding (8×16 atom favours N-alignment 2× over M-alignment) & \textasciitilde12\% & \textasciitilde4 TFLOPs & \textasciitilde0.5 \\ \hline
\textbf{GDDR6 channel-hash interaction} & Different (M, N) memory layouts produce different per-channel load distributions on 6 GDDR6 channels (residual after the other three) & - & - & \textasciitilde0.2~(residual) \\ \hline
\textbf{Total hardware-bound budget} & \textbf{~} & \textbf{~} & \textbf{~} & \textbf{5} \\ \hline
\end{longtable}

Table 15. Per-source distribution of the 5.0 TFLOPs / step hardware-bound roughness budget.

Per-source distribution of the 5.0 TFLOPs/step hardware-bound budget uses an impact-frequency times per-event-magnitude decomposition: for each mechanism we (a) define a measurable criterion under which the mechanism is ``active'' at a step transition, (b) compute the fraction of step transitions satisfying that criterion from the sweep data, and (c) compute the average \textbar dT\textbar{} at the active transitions minus the baseline \textbar dT\textbar{} at non-active transitions. The product gives the mechanism\textquotesingle s contribution to roughness. Table 15 summarizes the four contributions.

The wave-fill contribution combines two effects of the silicon-fixed C = 20 Xe-cores: (a) at small problems, fewer than 20 work-groups can be launched, leaving cores idle (the ramp component); (b) as the workload grows, W mod 20 varies across the sweep, producing tail-effect oscillation on top of the ramp. Both arise from the same silicon constraint and are computed as a single total variation of the wave-fill efficiency η\_W along the sweep, giving \textasciitilde3.0 TFLOPs / step at the optimal practical tile (T = 128).

Configuration-dependent overhead variation is bounded by the variance of the overhead surface across (M, N, K) configurations (Section 4.1, Figure 5 bottom-right). DPAS atom M:N asymmetry is bounded by the measured 2.5x N-cliff vs M-cliff asymmetry (Section 3.3) minus what padding removes. The GDDR6 channel-hash contribution is reported as the residual after the other three sources.

\subsection{Quantitative Root-Cause Summary}
Bringing together the per-cause magnitudes, Table 17 attributes the \textbf{16.8 TFLOPs / step initial landscape roughness} to its component sources, separating what software can remove from what is hardware-bound, and links each cause to the specific optimization or experiment that quantifies it.

Note on the decomposition: the entire 5 TFLOPs / step post-optimization residual is hardware-bound. Earlier in this paper we referred to a 2.0 TFLOPs / step "ideal-compute floor" as if it were a hardware-agnostic mathematical bound; this is a useful framing for the per-step roughness metric but the 2.0 value itself is a hardware consequence - specifically, the monotone-ramp climb that the silicon-fixed C = 20 Xe-cores impose between small-problem and saturated configurations. The decomposition below therefore classifies all 5 TFLOPs / step as hardware-bound, split between the wave-fill ramp (\textasciitilde2.0) and the per-feature oscillations on top (\textasciitilde3.0).

\begin{longtable}[]{!{\color{black}\vrule}
  >{\raggedright\arraybackslash}p{(\columnwidth - 6\tabcolsep) * \real{0.1534}}!{\color{black}\vrule}
  >{\raggedright\arraybackslash}p{(\columnwidth - 6\tabcolsep) * \real{0.3694}}!{\color{black}\vrule}
  >{\raggedright\arraybackslash}p{(\columnwidth - 6\tabcolsep) * \real{0.1711}}!{\color{black}\vrule}
  >{\raggedright\arraybackslash}p{(\columnwidth - 6\tabcolsep) * \real{0.3060}}!{\color{black}\vrule}}
\hhline{|-|-|-|-|}
\cellcolor{tblhdr}\begin{minipage}[b]{\linewidth}\raggedright
\textbf{Root cause}
\end{minipage} & \cellcolor{tblhdr}\begin{minipage}[b]{\linewidth}\raggedright
\textbf{Mechanism}
\end{minipage} & \cellcolor{tblhdr}\begin{minipage}[b]{\linewidth}\raggedright
\textbf{Magnitude (TFLOPs/step)}
\end{minipage} & \cellcolor{tblhdr}\begin{minipage}[b]{\linewidth}\raggedright
\textbf{Removed by}
\end{minipage} \\
\hline
\endhead

\endlastfoot
\multicolumn{4}{@{}>{\raggedright\arraybackslash}p{(\columnwidth - 6\tabcolsep) * \real{1.0000} + 6\tabcolsep}@{}}{%
\textbf{Software removable: \textasciitilde12 out of 16.8 reducible (71\% of original roughness)}} \\ \hline
\textbf{Coarse partial-tile waste} & Workload dimensions are not multiples of the workgroup tile size; the last WG along each dim wastes compute on padding. & \textasciitilde6.4 & Dynamic best-of-six tile selection (smaller tiles = less per-misalignment waste) \\ \hline
\textbf{Fine partial-tile waste at sub-128 N values} & Same mechanism as \#1 but at smaller scale: after picking a smaller tile, alignment still matters at 128-, 64-, 32-multiples. & \textasciitilde5.4 & DP padding (T1) - pads (M, N, K) up to a tile-aligned target with faster execution \\ \hline
\textbf{Pathological single-kernel shapes} & Different mechanism: shapes where no single kernel choice is fast (cache effects, wave-quantization tail, specific bad-N values); fixed by splitting into two sub-problems with better collective properties. & \textasciitilde0.4 & DP splitting (T2) - splits along K, N, or M into two sub-problems whose total time is lower \\ \hline
\multicolumn{4}{@{}>{\raggedright\arraybackslash}p{(\columnwidth - 6\tabcolsep) * \real{1.0000} + 6\tabcolsep}@{}}{%
\textbf{Hardware-bound: \textasciitilde5.0 residual after stack (29\% of original; 100\% of post-stack residual)}} \\ \hline
\textbf{Wave-fill effects on 20 Xe-cores} & Two causes:

a. Ramp from 20 fixed cores being unsaturated at small problems (\textasciitilde{} 2.0 TFLOPs/step)

b. Irregular wave-tail effects oscillate as workgroup count varies along the sweep (\textasciitilde0.8 TFLOPs/step oscillation) & \textasciitilde2.8 & Not removable; rooted in fixed core count vs. small-problem WG count. \\ \hline
\textbf{Per-kernel overhead variation} & Every kernel launch incurs \textasciitilde32\% of its time in fixed silicon costs (subgroup-barrier waits, register spills, GDDR6 latency that prefetch cannot hide). This overhead varies by \textasciitilde4 percentage points across (M, N, K) shapes - the variation, not the level, contributes to ruggedness. & \textasciitilde1.5 & Not removable; rooted in 256-GRF register file, sub-group barrier latency, GDDR6 latency-bound load semantics \\ \hline
\textbf{DPAS atom geometry drives M:N asymmetry} & N processed in 64-wide chunks, M in 32-wide chunks; N alignment matters 2.5x more than M & \textasciitilde0.5 & Not removable; architectural choice fixed in silicon \\ \hline
\textbf{GDDR6 memory channel imbalance} & 6 GDDR6 channels are loaded via a fixed address-interleaving hash. Different (M, N) memory layouts produce different per-channel load distributions; some shapes leave some channels idle while others overload, reducing effective bandwidth. & \textasciitilde0.2 & Not removable; memory subsystem fixed \\ \hline
\end{longtable}

\textbf{Table 16.} Quantitative attribution of the initial 16.8 TFLOPs / step landscape roughness on Intel Battlemage. Software techniques (best-of-six tile + DP pad + DP split) remove \textasciitilde12 (71\%) of the 16.8 TFLOPs / step total, leaving \textasciitilde5 of residual ruggedness that is dominated by wave-fill effects. Magnitudes for the hardware-bound sources are estimates inferred from the optimization stack\textquotesingle s residual roughness (5.0).

Of the 16.8 TFLOPs / step initial landscape roughness, \textbf{\textasciitilde12 (71\%) is software-removable} via our optimization stack (best-of-six tile selection contributes 6.4, DP padding contributes 5.4, DP splitting contributes 0.4); the remaining \textbf{\textasciitilde5 (29\%) is hardware-bound} and dominated by wave-fill effects across 20 fixed Xe-cores, configuration-dependent variation in the per-kernel base overhead and the DPAS atom\textquotesingle s M:N asymmetry.

\textbf{8.6 3D Aggregate View: Robustness of Conclusions Across the Full Sweep}

\includegraphics[width=\linewidth]{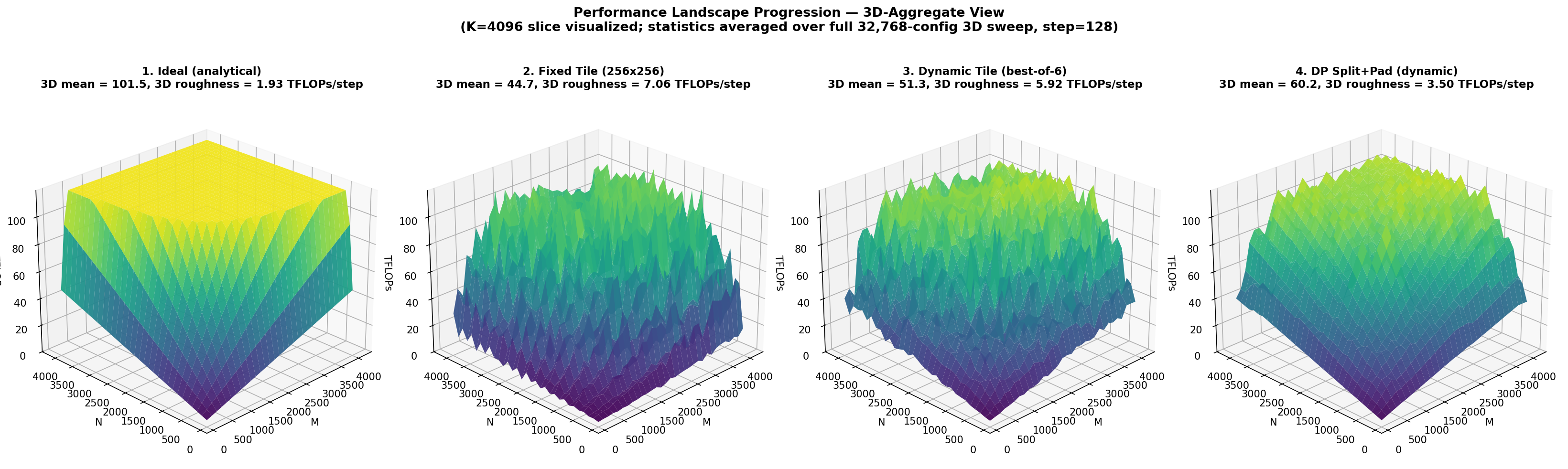}

Figure 21. Performance landscape progression at K = 4096 (visualized) with statistics aggregated over the full 32,768-configuration 3D sweep. Compare to Figure 1 (K = 4096 N-slice only).

All headline values in this paper (16.8 to 5.0 TFLOPs / step roughness reduction, 70\% improvement, 5 TFLOPs/step hardware-bound residual) are computed on the N-axis slice at the canonical K = 4096 configuration. This convention is appropriate because K = 4096 corresponds to the production-relevant GEMM regime (like transformer FFN layers, attention matrices) where ruggedness is most pronounced.

\begin{longtable}[]{!{\color{black}\vrule}
  >{\raggedright\arraybackslash}p{(\columnwidth - 4\tabcolsep) * \real{0.3121}}!{\color{black}\vrule}
  >{\raggedright\arraybackslash}p{(\columnwidth - 4\tabcolsep) * \real{0.3283}}!{\color{black}\vrule}
  >{\raggedright\arraybackslash}p{(\columnwidth - 4\tabcolsep) * \real{0.3595}}!{\color{black}\vrule}}
\hhline{|-|-|-|}
\cellcolor{tblhdr}\begin{minipage}[b]{\linewidth}\raggedright
\textbf{Stage}
\end{minipage} & \cellcolor{tblhdr}\begin{minipage}[b]{\linewidth}\raggedright
\textbf{Ruggedness at K=4096 slice}
\end{minipage} & \cellcolor{tblhdr}\begin{minipage}[b]{\linewidth}\raggedright
\textbf{Ruggedness across 3D aggregate}
\end{minipage} \\
\hline
\endhead

\endlastfoot
\textbf{Ideal} & 2.0 & 1.93 \\ \hline
\textbf{Fixed tile (256x256)} & 16.78 & 7..06 \\ \hline
\textbf{Dynamic tile} & 10.36 & 5.92 \\ \hline
\textbf{DP split + pad} & 5.00 & 3.50 \\ \hline
\textbf{Hardware bound residue} & 3 & 1.57 \\ \hline
\end{longtable}

Table 17. Comparison of K = 4096 slice values vs full-3D aggregate values (averaged across all 32 K-slices). The aggregate values are 0.4 to 0.7x the K = 4096 values because the K-axis is the smoothest axis (K-changes scale compute and memory equally). All values are in TFLOPs/step.

For completeness, Figure 21 shows the equivalent landscape progression with statistics aggregated over the full 32,768-configuration 3D sweep averaging N-axis, M-axis, and K-axis roughness across all slices. The aggregate values are uniformly smaller (factor of \textasciitilde2) because the K-axis is inherently smoother: K-changes scale both compute and memory equally, producing smaller per-step jumps than N or M changes. Despite the numerical compression, all qualitative conclusions hold: software optimization removes \textasciitilde50\% of initial roughness; the residual is dominated by silicon-quantization features decomposed in §8.5; the four hardware-bound source attribution is preserved (with proportionally smaller per-source magnitudes).

All conclusions of this paper hold under either reporting convention.

\section{GEMM Rugged Performance Landscape from First Principles}

This section shows the rugged landscape is analytic. We derive the entire optimal GEMM landscape on BMG - not merely its peak, but its ramp, its wave-quantization texture, its memory-channel behaviour, and the quantitative roughness metric itself - in closed form from a handful of datasheet integers: the Xe-core count (20), the DPAS shape, the channel count (6) and the bandwidth.

The ruggedness is not noise to be sampled; it is the interference pattern between a continuous demand - the problem size sweeping the grid and a discrete supply - work mapped onto 8×16×16 DPAS atoms and 16-wide sub-groups, packed into work-group tiles, issued in waves across the 20 Xe-cores, and fed by bytes interleaved over 6 GDDR6 channels - and interference patterns have closed forms.

Two consequences follow, and both matter to a practitioner. First, the optimal rugged landscape of an accelerator can be written down from its specification before any kernel exists, making ruggedness a design-time quantity for hardware/software co-design and competitive analysis rather than a post-hoc measurement. Second, because the closed form is the best any kernel could achieve, subtracting it from a measured landscape cleanly separates what is the silicon\textquotesingle s doing from what is the kernel\textquotesingle s; and because real kernels are rarely optimal, that difference doubles as a concrete, per-shape map of the optimization headroom still left in the software. We make the derivation and that separation precise below - and we are equally precise about the single ingredient that is not first-principles, the vendor\textquotesingle s proprietary memory-channel hash and exactly what it would add. We build three surfaces on the measured (M, N, K) grid, all as achieved throughput, TFLOPs = 2·M·N·K / t.

\subsection{The ideal ceiling and its ramp}
The ideal landscape is the compute ceiling.

The compute lower bound on time is \textbf{2MNK / P\textsubscript{peak}}, but a problem only earns the peak rate P\textsubscript{peak} once it has enough workgroups to occupy all C = 20 Xe-cores. With the default work-group tile T = 256 the work-group count is W = ⌈M/T⌉·⌈N/T⌉, and the ideal achieved throughput is P\textsubscript{peak} · min(1, W / C).

The ideal climbs from a small-problem minimum to the theoretical peak once one full wave fills the cores and is flat thereafter. It is the operation-independent reference - it encodes only the peak rate and the partial-wave fill on a fixed core count for the default tile. With the BF16 peak P\textsubscript{peak} = 108.2 TFLOP/s (20 Xe-cores × 8 DPAS units × 128 BF16 MAC/cycle × 2 FLOP/MAC × 2.644 GHz) it reproduces the ideal of §7.5 exactly: mean 97.2 TFLOPs, roughness 1.93 / step.

\subsection{The optimal: deriving ruggedness in closed form}
The leap of this section is that the rugged optimal - the best a real kernel could do, valleys and all - is the same ceiling perturbed by the hardware\textquotesingle s discrete units, and those perturbations are closed-form. For each (M, N, K) the optimal is the best, over the feasible work-group tiles T ∈ \{256, 128, 64\}, of the roofline

\textbf{t\_opt = min over T of max( t\_compute(T) , t\_memory(T) ), TFLOPs\_opt = 2MNK / t\_opt.}

\textbf{Compute - wave quantization across tiles}: For a tile T the kernel launches W\_T = ⌈M/T⌉·⌈N/T⌉ work-groups, run in ⌈W\_T/C⌉ discrete waves of C = 20 cores, so the compute efficiency is η\_wave(T) = W\_T / (⌈W\_T/C⌉·C) and t\_compute = 2MNK / (P\textsubscript{peak} · η\_wave).

A problem that needs 21 workgroups of a given tile pays for two full waves but populates only 21 of the 40 core-slots - 53\% efficiency. An optimal kernel escapes that tail by choosing the tile that best fills the cores: for a mid-size problem a finer tile yields more (smaller) workgroups and fills the 20 cores sooner than the default 256 tile. One consequence matters for reading Figure 22: because the ideal of §9.1 assumes the default 256 tile, the optimal - free to pick a finer tile - reaches near-peak earlier than the ideal ramp and therefore sits slightly above the ideal in the mid-size region. The ideal remains the operation-independent, default-tile reference; the optimal is its best-tile realisation.

\textbf{Memory GDDR6 channel roofline}: \textbf{t\_memory = MK·2/(BW·η\_A) + KN·2/(BW·η\_B) + MN·4/(BW·η\_D)}, with BW = 451 GB/s (192-bit GDDR6 at 2.4 GHz × 0.98). The per-tensor channel utilisation η is computed structurally, not fitted: GDDR6 stripes the address space across its 6 channels in 256 B blocks, and a kernel reads/writes 256-wide tile rows, so for a tensor of element size e (BF16 = 2 B for A, B; FP32 = 4 B for D) and leading dimension L (L = K for A, L = N for B and D) a row covers w = min(6, e) channels and its start block advances by s = (L·e / 256) mod 6 each row, hitting 6 / gcd(s, 6) distinct start channels, giving η = min(6, w · 6/gcd(s, 6)) / 6. Because only B and D carry L = N, η is N-asymmetric and dips (η\_B → 1/3) whenever N is a multiple of 768. Crucially this term only binds at low arithmetic intensity: for small or skinny shapes the optimal drops to the GDDR6 roofline, but at large M, N the intensity is far above the ridge (P\textsubscript{peak} / BW ≈ 240 FLOP/byte), so GDDR6 does not reduce the large-dimension plateau - it lowers only the small and skinny shapes.

Both terms are closed form, and every constant needed to reproduce Figure 22 is given above: P\textsubscript{peak} = 108.2 TFLOP/s, C = 20, the tile set \{256, 128, 64\}, BW = 451 GB/s, 6 channels at 256 B interleave, and the BF16/FP32 element sizes. The surface is a direct evaluation of these two formulas - one can recompute any (M, N, K) point by hand - not the output of a black-box tool. The derivation validates itself where it can be checked against the measurement it never used; the optimal bounds the measured throughput from above at 98\% of configurations; and its roughness reproduces the hardware-bound residual that extracts from the data by an entirely independent route.

\subsection{Result: optimal vs ideal vs actual}
\includegraphics[width=\linewidth]{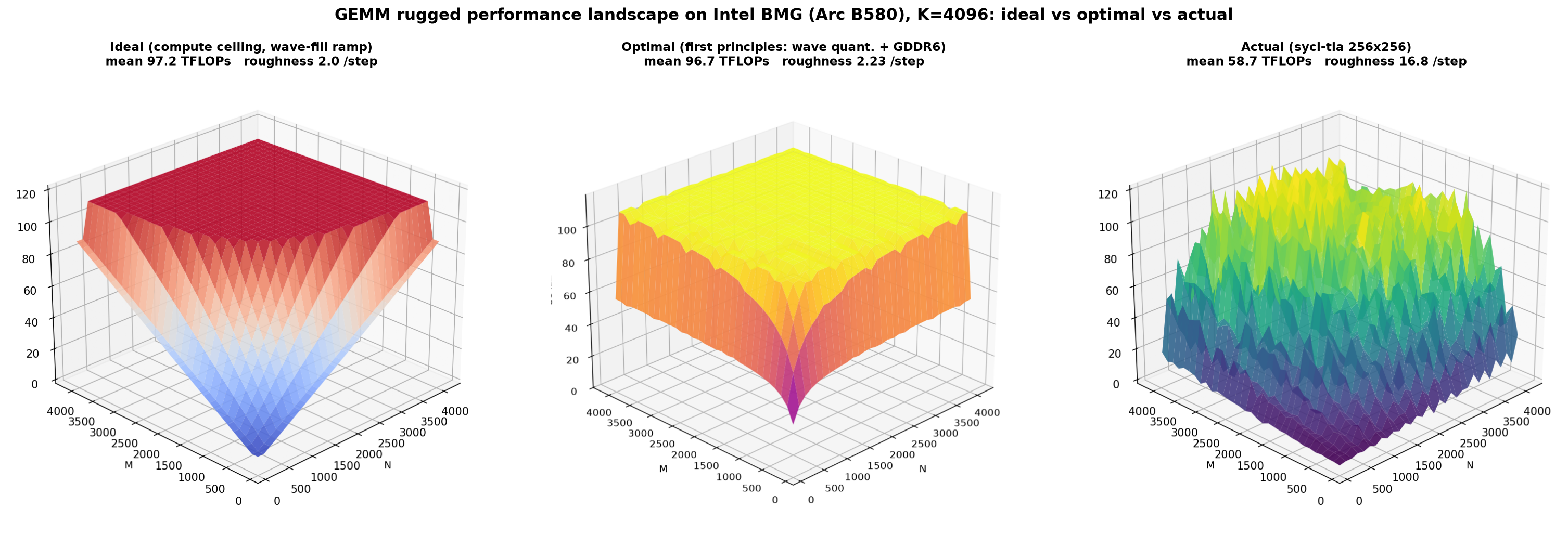}

Figure 22: GEMM landscape on BMG (Arc B580), K = 4096. Left: ideal compute ceiling (mean 97.2, roughness 2.0) - smooth. Centre: the first-principles optimal (mean 96.7, roughness 2.23) - near-peak across the large-dim plateau, dropping to the GDDR6 roofline only at small/skinny shapes, with wave-quant texture. Right: actual sycl-tla (mean 58.7, roughness 16.8).

\includegraphics[width=\linewidth]{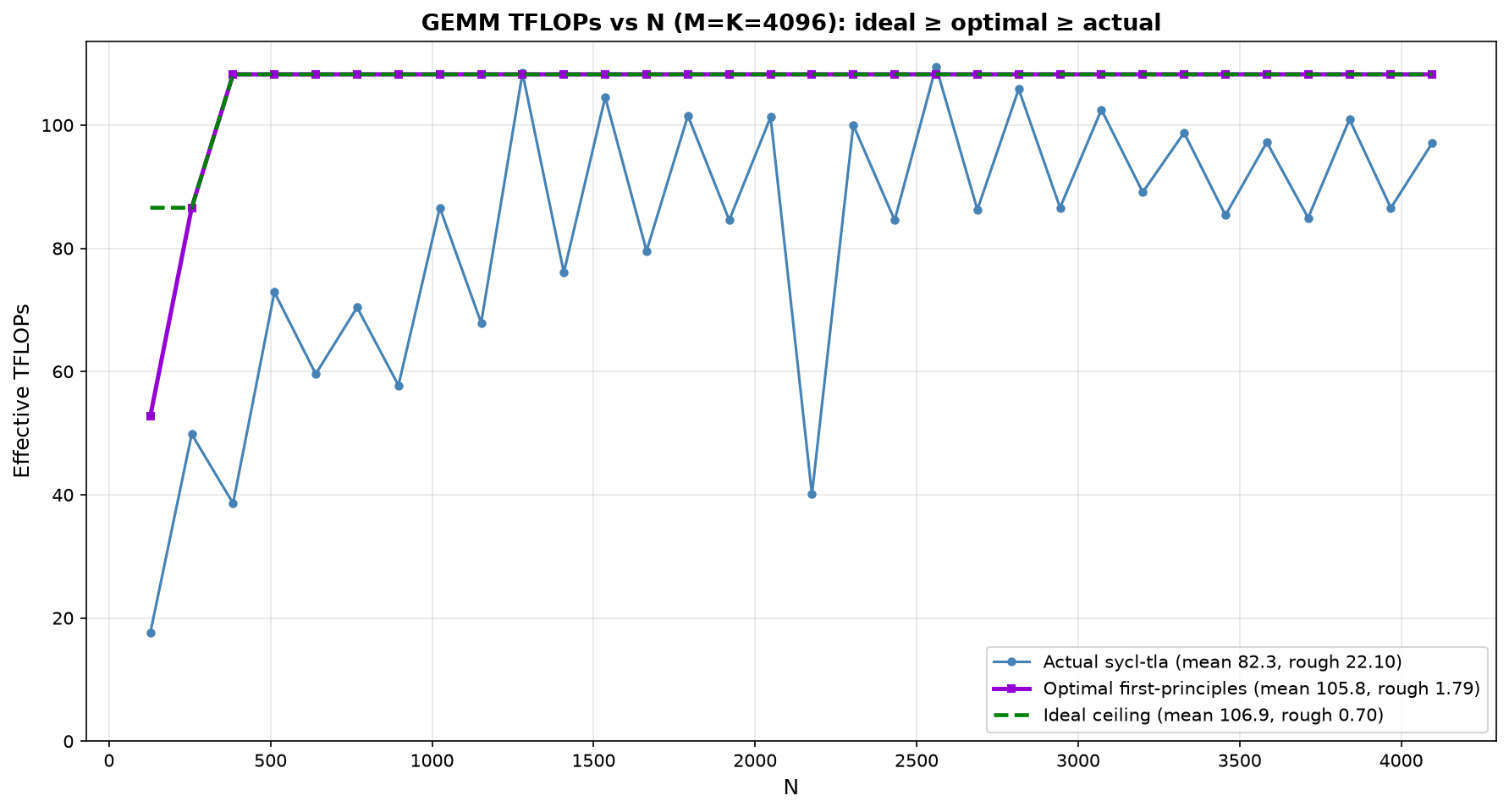}

Figure 23: Slice at M = K = 4096: the optimal tracks the ceiling near peak, while the actual kernel falls \textasciitilde30\% with the tile-boundary sawtooth.

\begin{longtable}[]{!{\color{black}\vrule}
  >{\raggedright\arraybackslash}p{(\columnwidth - 4\tabcolsep) * \real{0.3768}}!{\color{black}\vrule}
  >{\raggedright\arraybackslash}p{(\columnwidth - 4\tabcolsep) * \real{0.2898}}!{\color{black}\vrule}
  >{\raggedright\arraybackslash}p{(\columnwidth - 4\tabcolsep) * \real{0.3334}}!{\color{black}\vrule}}
\hhline{|-|-|-|}
\cellcolor{tblhdr}\begin{minipage}[b]{\linewidth}\raggedright
\textbf{Surface}
\end{minipage} & \cellcolor{tblhdr}\begin{minipage}[b]{\linewidth}\raggedright
\textbf{Mean TFLOPs}
\end{minipage} & \cellcolor{tblhdr}\begin{minipage}[b]{\linewidth}\raggedright
\textbf{Roughness / step}
\end{minipage} \\
\hline
\endhead

\endlastfoot
\textbf{Ideal landscape} & 97.2 & 1.93 \\ \hline
\textbf{Optimal (first principles)} & 96.7 & 2.23 \\ \hline
\textbf{Actual (sycl-tla, all optimizations)} & 76.3 & 3.50 \\ \hline
\end{longtable}

Table 18. Optimal vs ideal vs actual (\textbf{full-3D aggregates}).

The decomposition is mechanistic. Roughness climbs 1.93 to 2.23 to 3.50: the ideal is smooth by construction; the optimal carries the irreducible ruggedness - the closed-form wave-quantization and GDDR6-channel terms, landing on the \textasciitilde5 / step hardware residual §8 measured by a different method; and the actual stacks the software sawtooth on top.

In throughput, the optimal reaches a mean of 96.7 - it tracks the ideal ceiling and as it is free to pick the best tile, sits slightly above the default-tile ideal in the mid-size region while GDDR6 pulls it below the ideal only at small/skinny shapes - whereas the actual averages just 76.3. The gap optimal − actual (\textasciitilde20.4 TFLOPs of mean, \textasciitilde1.27 TFLOPs / step of roughness) is the software-removable loss the stack of §7 is built to recover. The \textbf{landscape is a fingerprint expressed as arithmetic}: the ideal is the mathematical bound, the optimal adds the silicon, the actual adds the kernel - so a vendor, an autotuner or a competitor can read off, per shape, how much headroom is available to win and how much the hardware already spent.

\subsection{The boundary of first principles}
Being explicit about what a datasheet cannot predict is as important as what it can. Two effects sit outside the closed form, for opposite reasons.

(i) Split-K - partitioning K to expose more parallelism - is a software feature that is not implemented in the sycl-tla kernel we measured; it is not a hardware limit, so this part of the optimal-vs-actual gap is unrealised software headroom (our DP optimizer independently elects split-K in \textasciitilde13\% of configurations).

(ii) The GDDR6 channel-conflict hash is a hardware feature, but a vendor-proprietary, undisclosed one: real controllers scramble physical addresses across channels and banks with a secret function to avoid pathological conflicts. Unlike split-K, it cannot be added to a first-principles projection at all, because the required information is not public - the projection captures only the documented 256B interleave, not the hidden scramble. Were the hash known, it would break the M ↔ N symmetry of the estimate - shifting the optimal peak off the square toward the rectangle the measurement prefers (3840 × 2048) - and add further conflict-driven roughness.

The optimal of Figure 22 is therefore a smooth lower bound on the true optimal ruggedness, and the gap to it marks the exact point where public specification ends and vendor-private microarchitecture begins.

\section{Kernel Optimality Levels: A Kardashev Scale for GPU Kernels}

\subsection{The issue of percent-of-peak}
Kernels are conventionally graded by peak efficiency η\_peak = A\_max / P - best measured throughput over datasheet peak or concerned input sizes. On our sweep the fixed-tile sycl-tla kernel scores η\_peak ≈ \textbf{0.95} (110.8 / 116.5 TFLOP/s), yet we demonstrated it attains a fraction of the achievable throughput across the (M, N, K) domain and runs several times rougher than the hardware forces.

η\_peak grades one favourable shape and is blind to both the rest of the surface and its predictability. Section §9 supplies the two missing instruments:

\begin{itemize}
\item
  the \textbf{first-principles optimal landscape O}
\item
  and the \textbf{roughness metric R}
\end{itemize}

which let us grade the whole surface on two axes: throughput attained and roughness above the irreducible hardware enforced floor.

\subsection{Two axes: how high and how smooth}
Over a domain D (the 32,768-point sweep), with first-principles optimal O(m, n, k) measured A(m, n, k), and roughness R (mean \textbar ΔTFLOPs\textbar{} along N at fixed M, K):

\begin{itemize}
\item
  \textbf{Attainment τ} = mean\_D A / mean\_D O ∈ {[}0, 1{]} - fraction of the achievable surface realized; τ = 1 implies no software-removable throughput loss remains.
\item
  \textbf{Roughness ratio ρ} = R(A) / R(O) ≥ 1 - multiples of the irreducible floor; ρ = 1 implies as predictable as the silicon allows.
\end{itemize}

Both are ratios against the GPU\textquotesingle s own analytic landscape, so they need no tuned reference kernel and cannot be gamed by cherry-picking a shape; η\_peak = A\_max / P is the degenerate one-shape case. As O omits the proprietary channel hash, R(O) underestimates the true floor so ρ is a conservative (upper-bound) grade.

\subsection{The optimality levels (L0 to L3)}
Just as the Kardashev scale types a civilization by the energy it commands, the Kernel Optimality Level (KOL) scales a kernel by how much of its GPU\textquotesingle s landscape it commands, gated by the weaker axis: L = min(T, S).

\begin{longtable}[]{!{\color{black}\vrule}
  >{\raggedright\arraybackslash}p{(\columnwidth - 8\tabcolsep) * \real{0.2448}}!{\color{black}\vrule}
  >{\raggedright\arraybackslash}p{(\columnwidth - 8\tabcolsep) * \real{0.2448}}!{\color{black}\vrule}
  >{\columncolor{tblhdr}}p{(\columnwidth - 8\tabcolsep) * \real{0.0320}}!{\color{black}\vrule}
  >{\raggedright\arraybackslash}p{(\columnwidth - 8\tabcolsep) * \real{0.1705}}!{\color{black}\vrule}
  >{\raggedright\arraybackslash}p{(\columnwidth - 8\tabcolsep) * \real{0.3079}}!{\color{black}\vrule}}
\hhline{|-|-|-|-|-|}
\cellcolor{tblhdr}\textbf{τ} & \cellcolor{tblhdr}\textbf{Throughput band} & & \cellcolor{tblhdr}\textbf{ρ} & \cellcolor{tblhdr}\textbf{Smoothness band} \\
\cline{1-2}\cline{4-5}
\textbf{\textless{} 0.50} & T0 peak-tuned & & \textgreater{} 3 & S0 rugged \\
\cline{1-2}\cline{4-5}
\textbf{0.50 to 0.75} & T1 partial & & 2 to 3 & S1 textured \\
\cline{1-2}\cline{4-5}
\textbf{0.75 to 0.95} & T2 near optimal & & 1.25 to 2 & S2 near floor \\
\cline{1-2}\cline{4-5}
\textbf{\textgreater= 0.95} & T3 optimal & & \textless= 1.25 & S3 at floor \\
\hhline{|-|-|-|-|-|}
\end{longtable}

Table 19. Different kernel optimality levels (KOLs) based on \textbf{τ} and \textbf{ρ}.

L3 (T3 ∧ S3) means the kernel is statistically indistinguishable from O in both throughput and roughness across D - the only remaining lever is different silicon.

An optimal kernel implementation must be landscape aware.

\subsection{Grading the sycl-tla GEMM kernel using KOL}
Evaluated on the K = 4096 slice against O and P = 116.5:

\begin{longtable}[]{!{\color{black}\vrule}
  >{\raggedright\arraybackslash}p{(\columnwidth - 8\tabcolsep) * \real{0.3329}}!{\color{black}\vrule}
  >{\raggedright\arraybackslash}p{(\columnwidth - 8\tabcolsep) * \real{0.1662}}!{\color{black}\vrule}
  >{\raggedright\arraybackslash}p{(\columnwidth - 8\tabcolsep) * \real{0.1477}}!{\color{black}\vrule}
  >{\raggedright\arraybackslash}p{(\columnwidth - 8\tabcolsep) * \real{0.1373}}!{\color{black}\vrule}
  >{\raggedright\arraybackslash}p{(\columnwidth - 8\tabcolsep) * \real{0.2159}}!{\color{black}\vrule}}
\hhline{|-|-|-|-|-|}
\cellcolor{tblhdr}\begin{minipage}[b]{\linewidth}\raggedright
\textbf{Kernel}
\end{minipage} & \cellcolor{tblhdr}\begin{minipage}[b]{\linewidth}\raggedright
\textbf{η\_peak}
\end{minipage} & \cellcolor{tblhdr}\begin{minipage}[b]{\linewidth}\raggedright
\textbf{τ}
\end{minipage} & \cellcolor{tblhdr}\begin{minipage}[b]{\linewidth}\raggedright
\textbf{ρ}
\end{minipage} & \cellcolor{tblhdr}\begin{minipage}[b]{\linewidth}\raggedright
\textbf{Level L}
\end{minipage} \\
\hline
\endhead

\endlastfoot
\textbf{First-principles optimal} & 0.962 & 1.00 & 1.00 & L3 (T3/S3) \\ \hline
\textbf{Fixed 256x256 (sycl-tla)} & 0.95 & 0.61 & 5.9 & L0 (T1/S0) \\ \hline
\textbf{Best-of-6 dynamic tile} & 0.962 & 0.71 & 3.7 & L0 (T1/S0) \\ \hline
\textbf{DP pad and split} & 0.962 & 0.79 & 1.8 & L2 (T2/S2) \\ \hline
\end{longtable}

Table 20. Grading of the 4 variants of sycl-tla GEMM kernel.

All three real kernels report η\_peak ≈ 0.95 to 0.962 - a straight A on the classical scale yet span L0 - L2, because peak efficiency cannot see the surface. DP pad and split (τ = 76.3 / 96.7 = 0.789) marginally clears the T1 to T2 boundary. The scale both credits and localizes the work: the DP stack climbs two levels to L2 (landscape-aware) and names the exact remaining gap to L3 - add \textbf{split-K} (unimplemented software feature; the DP already elects it in 13\% of shapes, Table 9) to push τ past 0.95, and trim R from 5.0 to ≤ 3.5 to reach S3.

To grade any kernel on any GPU, derive O from the datasheet, sweep A over D, and \textbf{report (τ, ρ, L)}. This replaces ``percent of peak'' with the operational question a deployment benefits from - what fraction of the achievable surface, how predictably, across every shape the workload will hit.

\section{Conclusions}

We proposed~\textbf{performance ruggedness analysis}~as an analytical framework complementary to the roofline model, treating the full GEMM performance landscape as the primary object of study and applied it to BF16 NN GEMM on Intel Battlemage (Arc B580). A 32,768-configuration sweep showed an initial landscape roughness of 16.8 TFLOPs / step against a hardware-determined floor of 5.0; a two-stage software stack (best-of-six dynamic tile selection plus a novel dynamic-programming padding-and-splitting optimizer) reduced roughness by~\textbf{70\%}~while raising mean TFLOPs by~\textbf{30\%}. The four-surface decomposition, randomized-order sweep methodology and cross-tile fine-N test collectively rule out cache set conflicts as an explanation for the residual periodic structure on this architecture, and instead attribute the remaining \textasciitilde29\% of original roughness to four hardware-bound sources: configuration-dependent variation in the 32\% per-kernel base overhead (rooted in the 256-GRF register file, subgroup barrier latency and GDDR6 latency-bound load semantics), wave quantization across 20 fixed Xe-cores, the 8x16x16 DPAS atom geometry\textquotesingle s M:N asymmetry and GDDR6 channel-hash interaction with~(M, N)~layouts.

The optimization stack we contribute is markedly simpler than autotuning: a one-time O(N⁴) precomputation on a 32,768-cell grid and the runtime decision is an O(1) table lookup over six precompiled tile variants. For workloads with shapes drawn from a known distribution, this stack achieves +30\% mean TFLOPs and −70\% landscape roughness on BMG GEMM without the cost-model or learned-dispatcher machinery production autotuners typically require.

The ruggedness-analysis framework itself has methodological implications independent of the optimizer: roofline-style scalar bounds reveal nothing about~how~achieved performance varies across input shapes, while ruggedness analysis decomposes that variation into software-removable and hardware-bound contributions usable for production service-level objective (SLO) design and competitive benchmarking.

The performance landscape is a fingerprint of the operation being profiled, software stack and hardware and can be leveraged for comparing the competitiveness of different software/hardware stacks.

\section*{References}
\small
\begin{enumerate}[leftmargin=2.2em,itemsep=1pt,topsep=2pt]
\def\labelenumi{[\arabic{enumi}]}
\item
  S. Williams, A. Waterman, and D. Patterson, "Roofline: an insightful visual performance model for multicore architectures," Communications of the ACM, vol. 52, no. 4, pp. 65--76, Apr. 2009.
\item
  H. Li, Z. Xu, G. Taylor, C. Studer, and T. Goldstein, "Visualizing the loss landscape of neural nets," in Advances in Neural Information Processing Systems (NeurIPS), vol. 31, Montréal, Canada, Dec. 2018, pp. 6389--6399.
\item
  Intel Corporation, "Intel Arc B580 Graphics Card Specifications," 2024. {[}Online{]}. Available: \url{https://www.intel.com/content/www/us/en/products/sku/241598/intel-arc-b580-graphics/specifications.html} (accessed May 14, 2026).
\item
  Intel Corporation, "sycl-tla: SYCL port of CUTLASS Templates for Linear Algebra Subroutines," GitHub, 2024 (commit be58860e). {[}Online{]}. Available: \url{https://github.com/intel/sycl-tla} (accessed May 14, 2026).
\item
  NVIDIA Corporation, "CUTLASS: CUDA Templates for Linear Algebra Subroutines," GitHub, 2017--present. {[}Online{]}. Available: \url{https://github.com/NVIDIA/cutlass} (accessed May 14, 2026).
\item
  JEDEC Solid State Technology Association, JESD250C: Graphics Double Data Rate 6 (GDDR6) SGRAM Standard, Arlington, VA, USA, Aug. 2017.
\item
  J. D. McCalpin, "Memory bandwidth and machine balance in current high performance computers," IEEE Computer Society TCCA Newsletter, pp. 19--25, Dec. 1995.
\item
  V. Volkov, "Better performance at lower occupancy," presented at the GPU Technology Conference (GTC), San Jose, CA, USA, Sep. 2010. {[}Online{]}. Available: \url{https://www.nvidia.com/content/GTC-2010/pdfs/2238_GTC2010.pdf} (accessed May 14, 2026).
\item
  Intel Corporation, Intel VTune Profiler User Guide, 2024. {[}Online{]}. Available: \url{https://www.intel.com/content/www/us/en/develop/documentation/vtune-help/top.html} (accessed May 14, 2026).
\end{enumerate}

\end{document}